\documentclass{aa} % for two column printer version
\pdfoutput=1

\usepackage{xcolor}
\usepackage[normalem]{ulem}
\usepackage{cancel}
\usepackage{tikz}

\usepackage{multicol}
\usepackage{graphicx}
\usepackage{wrapfig}
\usepackage{txfonts}
\usepackage{diagbox}

\usepackage{color}

\usepackage{mathtools}

\usepackage{ulem}

\DeclarePairedDelimiter\floor{\lfloor}{\rfloor}

\newcommand{\dt}[1]{\frac{\partial{#1}}{\partial{t}}}

\newcommand{\coma}{\hspace{3pt},}
\newcommand{\punto}{\hspace{3pt}.}
\newcommand{\pcoma}{\hspace{3pt};}

\newcommand\nth{\textsuperscript{th}\xspace} %\th is taken already
\newcommand\mancha{\textsc{Mancha3D~}}
\newcommand\phantoms{\textsc{Phantom-spmhd~}}
\newcommand\muram{\textsc{MUR}a\textsc{M~}}

\newcommand\bifrost{\textsl{Bifrost~}}

\begin{document} 

    \title{MHDSTS: a new explicit numerical scheme for simulations of partially ionised solar plasma.}

    \author{P. A.~Gonz\'alez-Morales\inst{1,2}
        \and
            E.~Khomenko\inst{1,2}
        \and
            T. P.~Downes\inst{3}
        \and
            A.~de Vicente\inst{1,2}
    }

    \institute{Instituto de Astrof\'isica de Canarias (IAC), E-38205 La Laguna, Tenerife, Spain,
  	    \and
  		    Universidad de La Laguna (ULL), Dpto. Astrof\'isica, E-38206 La Laguna, Tenerife, Spain,
            %\email{pagm@iac.es}\\	
            %\email{khomenko@iac.es}\\
            %\email{angelv@iac.es}
	    \and
            Centre for Astrophysics \& Relativity, School of Mathematical 
				Sciences, Dublin City University (DCU), Ireland
            %\email{turlough.downes@dcu.ie}
        }

  \date{}

% \abstract{}{}{}{}{} 
% 5 {} token are mandatory

 \abstract{The interaction of plasma with magnetic field in the partially ionised solar atmosphere is frequently modelled via a single-fluid approximation, which is valid for the case of a strongly coupled collisional media, such as solar photosphere and low chromosphere. Under the single-fluid formalism the main non-ideal effects are described by a series of extra terms in the generalised induction equation and in the energy conservation equation. These effects are: Ohmic diffusion, ambipolar diffusion, the Hall effect, and the Biermann battery effect. From the point of view of the numerical solution of the single-fluid equations, when ambipolar diffusion or Hall effects dominate can introduce severe restrictions on the integration time step and can compromise the stability of the numerical scheme. In this paper we introduce two numerical schemes to overcome those limitations. The first of them is known as \emph{Super Time-Stepping} (STS) and it is designed to overcome the limitations imposed when the ambipolar diffusion term is dominant. The second scheme is called the \emph{Hall Diffusion Scheme} (HDS) and it is used when the Hall term becomes dominant. These two numerical techniques can be used together by applying Strang operator splitting. This paper describes the implementation of the STS and HDS schemes in the single-fluid code \textsc{Mancha3D}. The validation for each of these schemes is provided by comparing the analytical solution with the numerical one for a suite of numerical tests.}

 % context heading (optional)
 % {} leave it empty if necessary  
 % {context}
 % aims heading (mandatory)
 % {aims}
 % methods heading (mandatory)
 % {methods}
 % results heading (mandatory)
 % {results}
 % conclusions heading (optional), leave it empty if necessary 
 % {conclusions}


  \titlerunning{New Explicit Numerical Scheme}
  \authorrunning{P.A.~Gonz\'alez-Morales et al.}
  \maketitle

%%%%%%%%%%%%%%%%%%%%%%%%%%%%%%%%%%%%%%%%%%%%%%%%%%%%%%%%%
\section{Introduction}\label{sec:intro}
%%%%%%%%%%%%%%%%%%%%%%%%%%%%%%%%%%%%%%%%%%%%%%%%%%%%%%%%%

Magnetic fields play an important role in the dynamics of many physical systems. The ideal magnetohydrodynamic (MHD) theory has been an excellent tool to understand the physical interaction between the magnetic field and the plasma in the solar atmosphere, with a long list of examples of successful modelling  of a number of phenomena \citep[e.g.,][]{2000A&A...359..729A, 2005ESASP.596E..65S, Vogler:2005hp,2006ApJ...642.1246S, 2006ApJ...653..739K, 2009LRSP....6....2N}.
Nevertheless, the assumptions underlying ideal MHD theory are not always valid. In the conditions of solar photosphere and low chromosphere, the presence of neutrals together with the weakening of collisional coupling due to the density fall off with height leads to the invalidation of the assumptions underpinning the MHD approach and produces a series of non-ideal effects. These non-ideal effects have been actively studied during recent years, see for example \citet{2003SoPh..214..241K, 2004A&A...422.1073K, 2007A&A...461..731F, 2008MNRAS.385.2269P, Vranjes:2008kh, 2009ApJ...699.1553S, 2010A&A...512A..28S, 2011A&A...529A..82Z, 2012ApJ...747...87K, 2012ApJ...753..161M}. The importance of the non-ideal effects can be evaluated by examining the ratio between the cyclotron frequency and the collisional frequencies. In the photosphere, the cyclotron frequency can exceed the collisional one in strongly magnetised areas such as flux tubes and sunspots, while in the chromosphere this happens over most of its volume even in regions with a weaker magnetic field \citep{2014PhPl...21i2901K}. It can be expected that in such conditions the neutral and charged particles behave in a different way, with direct repercussions on the plasma dynamics and energy exchange. A recent review of the behaviour of partially ionised plasmas in different astrophysical conditions is provided by \citet{2017arXiv170707975B}.

Mathematically, the treatment of partially ionised plasmas depends on the degree of collisional coupling. The most widely used formalisms are either a two-fluid or single-fluid approach, see for example the recent discussion in \citet{2014PhPl...21i2901K} and a review by \citet{2017arXiv170707975B}. The two-fluid formalism, including the treatment of ionisation-recombination terms was recently provided by \citet{2012PhPl...19g2508M}. While the single-fluid formalism yields a less detailed description, it is numerically more efficient for strongly stratified solar plasma, and therefore it has been adopted in several numerical codes that aim at realistic simulations of the solar atmosphere such as \muram \citep{Vogler:2005hp}, \bifrost \citep{2011A&A...531A.154G} and \mancha \citep{2010ApJ...719..357F}, the latter being the one used in the current study. The single-fluid formalism has an advantage since numerically problematic collisional terms (both elastic collisions and ionisation-recombination) do not appear explicitly in the equations, greatly simplifying the modelling. For the range of collisional frequencies typical for the solar photosphere and chromosphere, these collisional terms require extremely short time steps, unsuitable for most frequently used explicit numerical codes. In the solar case, {the single-fluid formalism} can safely be used for the modelling of the dense and collisional photosphere and low chromosphere, for the typical time scales of interest \citep{2011A&A...529A..82Z}. For both the single and two-fluid formalisms the generalised Ohm's law \citep{Cowling:1957ul} has to be used to describe the behavior of electric currents. The generalised Ohm's law has a series of additional terms derived from the presence of neutrals, while other standard terms have slightly modified expressions \citep{2014PhPl...21i2901K}. As a consequence, there appear additional terms in the induction and in the energy conservation equations. The leading non-ideal terms in the single-fluid formalism are: the Ohmic diffusion, the ambipolar diffusion, Hall term, and Biermann battery term.

In an astrophysical context, ambipolar diffusion is envisaged as a process of diffusion suffered by the magnetic field due to collisions between neutral and charged particles, the latter being frozen into the field. Ambipolar diffusion helps converting magnetic energy into thermal energy on scales that are typically much faster than classical Ohmic diffusion. For the Sun, it has been demonstrated that the ambipolar diffusion is orders of magnitude larger than Ohmic diffusion in the chromospheric layers, therefore greatly contributing to chromospheric heating via Joule dissipation \citep{2012ASPC..463..281K,Priest:2014we}. In the two-fluid treatment, the energy equation for the charged fluid also has a Joule heating term, similar to the single-fluid case, and the expression for the electric field comes from the generalised Ohm's law \citep[see, e.g.,][]{2014PhPl...21i2901K, 2017arXiv170707975B}. It is interesting to note, however, that ambipolar diffusion term is not explicitly present in the two-fluid Ohm's law, and in the energy conservation equation in the two-fluid treatment, and the corresponding effect is expressed via explicit ion-neutral collisional terms.

The Hall effect appears when ions decouple from the magnetic field while electrons remain attached to it. In a fully ionised plasma ions decouple from the magnetic field due to the difference between the inertia of ions and electrons, and this decoupling typically happens for processes at high frequencies. In weakly ionised plasmas, however, the ions can become detached from the magnetic field due to collisions with neutrals, leading to a similar effect. The spatial and temporal scales of the Hall effect in partially ionised plasmas are different from those in fully ionised plasmas, but the dynamics are similar in both cases. The Hall term is not dissipative, so it does not contribute to the heating of the environment, however it does redistribute magnetic energy in the system. Therefore, the Hall effect has an important role at many scales, from laboratory to astrophysical plasmas, while ambipolar diffusion is typically only important at low ionisation fractions \citep{Parker:2007wf, 2008MNRAS.385.2269P, Priest:2014we, 2017arXiv170707975B}.

The Biermann battery effect is somewhat different to the two non-ideal effects discussed above. Its magnitude in the solar case is rather small and it is usually ignored as it is not deemed important. But recently \citet{2017A&A...604A..66K} have shown that this term is able to seed the magnetic field for the local solar dynamo. In the present paper we will consider this term for completeness. 

From a numerical point of view, the battery term just introduces a small source term into the hyperbolic system of equations of single-fluid MHD. As long as this source term is small in some sense it is not necessary for either accuracy or stability reasons to introduce a time step restriction based on it. However, the ambipolar term is diffusive (i.e., parabolic), and the Hall term is dispersive. These two latter terms introduce important restrictions into the integration time step and into the stability of explicit numerical schemes, which are frequently used in hydrodynamic codes solving hyperbolic equations.

This paper presents a new numerical scheme aiming to overcome these restrictions. This scheme is implemented as separate modules in the existing single-fluid radiative MHD code \mancha \citep{2006ApJ...653..739K, 2012ApJ...747...87K, 2010ApJ...719..357F, Vitas:AA2016}. The proposed scheme is relatively easy to add into an existing explicit numerical scheme, which  allows one to keep all the benefits and flexibility of the explicit approach. We also provide a suite of numerical tests to verify the newly implemented modules and to check their order of accuracy.

This paper is organised as follows. In Section \ref{sec:eq} we show the
set of equations corresponding to a single fluid description with a
generalised Ohm's law. In Section \ref{sec:num_sch} we present very
briefly our single-fluid code \mancha and the details of the new
numerical scheme introduced here. In Section \ref{sec:test} we show
tests demonstrating the feasibility and reliability of the new scheme
introduced, while in Section \ref{sec:sum} a brief summary is presented.

%%%%%%%%%%%%%%%%%%%%%%%%%%%%%%%%%%%%%%%%%%%%%%%%%%%%%%%%%
\section{Equations}\label{sec:eq}
%%%%%%%%%%%%%%%%%%%%%%%%%%%%%%%%%%%%%%%%%%%%%%%%%%%%%%%%%
As discussed in the Introduction, the peculiarity of solar plasma in the photosphere and in the low chromosphere is that, while being only weakly ionised, it is still very dense and, therefore, collisions play an important role in coupling all plasma and neutral components. Since typically the scales of interest are significantly larger than collisional scales \citep[see][]{ 2014PhPl...21i2901K}, the most efficient approach for the mathematical treatment of such plasma is a single-fluid quasi-MHD formalism. The derivation of conservation equations and the generalised induction equation for such a case can be consulted elsewhere \citep[see e.g.,][]{2017arXiv170707975B}.
Following the derivation of the equations from \citet{2004fopp.book.....B}, and using a specific definition for the macroscopic variables, see Appendix D of \citet{2014PhPl...21i2901K}, the non-ideal effects are encoded into several terms in the induction and total energy conservation equations. The set of equations is formed by the continuity equation,
%%%
\begin{equation}\label{eq:cont}
    \dt{\rho}+\nabla\cdot(\rho \mathbf{v})=0\coma 
\end{equation}
%%%
the momentum conservation equation,
%%%
\begin{equation}\label{eq:mom}
    \frac{\partial(\rho\mathbf{v})}{\partial t} + \nabla \cdot \Bigg[\rho\mathbf{vv}+\Bigg(p+\frac{\mathbf{B}^2}{2\mu_0}\Bigg)\mathbf{I}-\frac{\mathbf{BB}}{\mu_0}\Bigg]= \rho \mathbf{g}+\mathbf{S}(t) \coma
\end{equation}
%%%
and the induction equation,
%%%
\begin{equation}\label{eq:induc}
    \begin{aligned}
        \frac{\partial \mathbf{B}}{\partial t} = {} & \nabla \times \Bigg[(\mathbf{v}\times\mathbf{B})-\eta\mu_0\mathbf{J}-\eta_H\frac{\mu_0}{|B|}(\mathbf{J}\times\mathbf{B}) + {}  \\ & + \eta_A\frac{\mu_0}{|B|^2}\left[(\mathbf{J}\times\mathbf{B})\times\mathbf{B}\right] + \eta_B\frac{\mu_0}{|B|}\nabla p_e \Bigg] \coma%- \eta_P(\mathbf{G}\times\mathbf{B}) \Bigg] \coma
    \end{aligned}
\end{equation}
%%%
where we retained the convective, Ohmic, Hall, Ambipolar, and Biermann battery terms on the right hand side. The coefficients multiplying these terms are, respectively, the Ohmic ($\eta$), Hall ($\eta_H$), ambipolar ($\eta_A$), and battery ($\eta_B$) coefficients, and all have the same units of diffusivity ($l^2t^{-1}$, i.e., $m^2s^{-1}$ in SI). Those coefficients are defined later on, in equations (\ref{eq:ohm}), (\ref{eq:hall_bat}), and (\ref{eq:ambi}).
The total energy conservation equation,
%%%
\begin{equation}\label{eq:ener}
    \begin{aligned}
        \dt{e}{} & + \nabla\cdot\left[ \frac{\rho \mathbf{v}^2}{2}\mathbf{v}+\frac{\gamma p}{\gamma-1}\mathbf{v} + \frac{1}{\mu_0} \nabla\cdot\left[\mathbf{B}\times\left(\mathbf{v}\times\mathbf{B}\right)\right] \right]= \rho\mathbf{v}\cramped{\cdot}\mathbf{g}+ {}\\ &+ \nabla\cdot\left[\mathbf{B}\times\left(\eta+\eta_A\right)\mathbf{J}_\perp\right] + \nabla\cdot\left[\eta_B\frac{\nabla p_e\times\mathbf{B}}{|B|}\right]+ Q_\textrm{rad} \coma
    \end{aligned}
\end{equation}
%%%
is written in terms of the total energy density per volume unit $e$, which is the sum of the kinetic, internal and magnetic energies,
%%%
\begin{equation}\label{eq:totener}
    e=\frac{1}{2}\rho \mathbf{v}^2+\frac{p}{\gamma-1}+\frac{\mathbf{B}^2}{2\mu_0}\coma
\end{equation}
%%%
\noindent The electric currents are defined as
%%%%
\begin{equation}
    \mu_0\mathbf{J}=\nabla\times\mathbf{B} \coma
\end{equation}
\begin{equation}
    \mathbf{J}_\perp=-\frac{(\mathbf{J}\times\mathbf{B})\times\mathbf{B}}{|B|^2} \coma
\end{equation}
%%%
and Gauss's law for magnetism is
%%%
\begin{equation}\label{eq:gauss}
    \nabla\cdot{\bf B}=0\punto
\end{equation}
%%%
To close the system, the equation of state is used.

Equations (\ref{eq:mom}) and (\ref{eq:ener}) have two extra terms: an external time dependent force $\mathbf{S}(t)$ describing an arbitrary external perturbation, and a source term $Q_\textrm{rad}$ to take into account radiative energy exchange. These terms will be neglected in this work since the implementation of the numerical scheme introduced below is independent of the presence of these terms.\\
The coefficients multiplying the non-ideal terms in the induction and energy equations are defined as:
%%% 
\begin{equation}\label{eq:ohm}
    \eta = \frac{m_e(\nu_{ei}+\nu_{en})}{e^2n_e\mu_0} \coma
\end{equation}
%%%
\begin{equation}\label{eq:hall_bat}
    \eta_H=\eta_B=\frac{|B|}{en_e\mu_0} \coma
\end{equation}
%%%
\begin{equation}\label{eq:ambi}
    \eta_A=\frac{(\rho_n/\rho)^2|B|^2}{(\rho_i\nu_{in}+\rho_e\nu_{en})\mu_0} \coma
\end{equation}
%%%
The expressions for the collisional frequencies between ions and neutrals ($\nu_{in}$) and  electrons with neutrals ($\nu_{en}$) are taken from  \cite{1956pfig.book.....S}. For the collisions between electrons and ions ($\nu_{ei}$) we use the expression from \cite{1965RvPP....1..205B}: 
%%%
\begin{equation}
    \nu_{in}=n_n\sqrt{\frac{8k_BT}{\pi m_{in}}}\sigma_{in} \coma
\end{equation}
%%%
\begin{equation}
    \nu_{en}=n_n\sqrt{\frac{8k_BT}{\pi m_{en}}}\sigma_{en} \coma
\end{equation}
%%%
\begin{equation}
    \nu_{ei}=\frac{n_ee^4\ln{\Lambda}}{3\epsilon_0^2m_e^2}\left(\frac{m_e}{2\pi k_BT}\right)^{3/2} \coma
\end{equation}
%%%
where $m_{in}=m_im_n/(m_i+m_n)$ and $m_{en}=m_em_n/(m_e+m_n)$ are the reduced masses. The cross sections for a weakly ionised plasma assuming elastic collisions between solid spheres are $\sigma_{in}=5\times10^{-19}$m$^2$ and $\sigma_{en}=10^{-19}$m$^2$ \citep{Vranjes:2013dh, Huba:2016wq}. The Coulomb logarithm, $\ln\Lambda$, is  defined as
%%%
\begin{equation}
    \Lambda=\frac{12\pi(\epsilon_0k_BT)^{3/2}}{n_e^{1/2}e^3} \pcoma
\end{equation}
%%%
\begin{equation}
    \ln{\Lambda}\approx16.33-\frac{1}{2}\ln{\left(n_e[\mathrm{m}^{-3}]\right)}+\frac{3}{4}\ln{\left(T[\mathrm{K}] \right)} \punto
\end{equation}
%%%
The equations above assume charge neutrality ($n_e=n_i$) and negligible electron mass ($m_e/m_i\approx 0$). 
A complete derivation of these equations and a discussion of the relative importance of the non-ideal terms along the solar atmosphere can be found in \citet{2014PhPl...21i2901K}.

%%%%%%%%%%%%%%%%%%%%%%%%%%%%%%%%%%%%%%%%%%%%%%%%%%%%%%%%%
\section{Description and implementation of the new numerical scheme}\label{sec:num_sch}
%%%%%%%%%%%%%%%%%%%%%%%%%%%%%%%%%%%%%%%%%%%%%%%%%%%%%%%%%
%Before describing the MHDSTS\footnote{MHDSTS is the abbreviation for \textsc{Mhd}, \textsc{HDS} and \textsc{sTS} } scheme, we first briefly introduce our single-fluid \mancha code, which as mentioned in Section \ref{sec:intro}, is the code which we have used to implement and validate the new scheme.
\subsection{\mancha code}\label{sec:mancha}
Before describing the MHDSTS\footnote{MHDSTS is the abbreviation for \textsc{Mhd}, \textsc{HDS} and \textsc{sTS} } scheme, we first briefly introduce our single-fluid \mancha code. The current implementation of this numerical code is a parallel tridimensional extension of the code developed by \citet{2006ApJ...653..739K} and solves the non-linear 3D MHD equations for the perturbations written in conservative form \citep{2010ApJ...719..357F}.

To avoid high frequency numerical noise and improve the numerical stability some artificial diffusion terms are added instead of their physical counterpart following the work of \cite{Vogler:2005hp}, so each physical diffusivity has its own contribution formed by a constant contribution, a hyperdiffusivity part and a shock-resolving term. On the other hand, because in some simulations a high diffusivity can modify the perturbations' amplitude (especially when solving waves) the code can perform an additional filtering to damp small wavelengths using the method described by \cite{2007ApJ...666..547P}. In this way, the artificial diffusivity remains low and the numerical noise is avoided. For spatial integration \mancha uses a centred 6\nth order accurate derivative, following  \cite{Nordlund:1995un}. To advance in time it uses an explicit multi-step Runge-Kutta (RK) scheme, accurate to 4\nth order \citep{Vogler:2005hp, 2006ApJ...653..739K, 2010ApJ...719..357F}. To prevent wave reflection at the boundaries, especially when working with complex magnetic field configurations, the code has implemented a Perfectly Matched Layer (PML; \citealt{1994JCoPh.114..185B, Berenger1996363, 1996JCoPh.129..201H, Hesthaven1998129, 2001JCoPh.173..455H, 2009ApJ...694..573P}). It also has the possibility of using some Adaptive Mesh Refinement (AMR) method  \citep{1989JCoPh..82...64B} to follow the evolution of fine structures.

More recently, a new radiative transfer module has been added to solve the radiative transfer equations using the short characteristic method, being also possible to choose between an ideal equation of state or a realistic one in thermodynamical equilibrium based on precalculated tables, which include the ionisation equilibrium solution and the electron pressure needed to calculate the non-ideal terms, see \citet{2012ApJ...747...87K, Vitas:AA2016} for details.

The purely MHD version of the code has been verified against standard numerical tests by  \citet{2010ApJ...719..357F} and the code has been used so far for a wide range of numerical experiments (both ideal and realistic), such as modelling of wave propagation, instabilities, and magneto-convection \citep[e.g.,][]{2010ApJ...719..357F, 2012ApJ...747...87K, Felipe:2013el, Luna:2016fn, Felipe:2016bz,2016A&A...590L...3S,2017A&A...604A..66K}.

\bigskip

At this point, it is convenient to briefly explain the time integration method used by \mancha as we shall refer to it in the following sections. In the single-fluid approximation, the system of equations written in conservative form can be schematically represented as:
%%%
\begin{equation}\label{eq:conser}
    \dt{\mathbf{u}} = \mathcal{R}(\mathbf{u}) = -\nabla \mathbf{F}(\mathbf{u}) + \mathbf{S}(\mathbf{u}) + \mathbf{G}(\mathbf{u}) \coma
\end{equation}
%%%
where $\mathcal{R}(\mathbf{u})$ represents an operator formed by the sum of the spatial derivatives of fluxes $\mathbf{F}(\mathbf{u})$, the source terms, $\mathbf{S}(\mathbf{u})$, and the non-ideal terms (Ohmic, Ambipolar, Hall, and Battery), $\mathbf{G}(\mathbf{u})$. The vector $\mathbf{u}$ is the set of variables in the equations: $[\rho,\rho\mathbf{v},\mathbf{B},e](\mathbf{r},t)$.

The next time step is calculated using an explicit RK scheme that can be written in a compact form as: %\azul{[cuando k es 1 uk es un. Lo explicitamos? Quitar la indentacion de la primera]} 
%%%
\begin{subequations}\label{eq:rkm}
    \begin{alignat}{2}
%\mathbf{u}^{(0)} & = \mathbf{u}^{(n)} \coma\\
        \mathbf{u}^{(k)}  & = \mathbf{u}^{(n)}+\alpha_k \Delta t \mathcal{R} \left(\mathbf{u}^{(k-1)}\right)\coma \quad k=1,\dots,m\coma \\
        \mathbf{u}^{(n+1)} & = \mathbf{u}^{(m)}\coma 
    \end{alignat}
\end{subequations}
%%%
where $\mathbf{u}^{(n+1)}$ corresponds to the solution at $t_{n+1}=t_n+\Delta t$, with $\Delta t$ given by the Courant, Friedrichs and Lewy (CFL) condition. The order of the RK scheme is given by the chosen $m \le 4$, and the coefficients $\alpha_k$ can be found in \citet[Table 6]{1992CANM....8..761V} or easily computed using the expression (\ref{eq:ak}).
\begin{equation}\label{eq:ak}
    \alpha_k = \frac{1}{m+1-k} \punto
\end{equation}
\mancha uses a fourth order RK scheme to solve the non-ideal single-fluid MHD equations, which using this notation corresponds to $m=4$ in eqs. (\ref{eq:rkm}).
%%%%%%%%%%%%%%%%%%%%%%
\subsection{The new MHDSTS scheme}

As mentioned  in Section \ref{sec:intro}, the non-ideal MHD terms are numerically problematic. In the solar atmosphere the value of the ambipolar diffusion coefficient ($\eta_A$) strongly increases with height and becomes larger than the classical Ohmic diffusion ($\eta$) by several orders of magnitude \citep{2014PhPl...21i2901K}. When the ambipolar term dominates the equations, the system turns from hyperbolic to parabolic, and very small integration time steps are required in order to fulfil the CFL condition. To avoid such small time steps, the proposed MHDSTS scheme uses the technique known as Super Time-Stepping (STS). This numerical technique was proposed by \citet{Gentzsch1980} to speed up explicit time-stepping schemes for parabolic problems. Later on, \citet{Alexiades:1995wj} and \citet{Alexiades:1996vj} presented a variant of this technique, which is used by \citet{2006MNRAS.366.1329O, 2007MNRAS.376.1648O} to overcome the limitations caused by a large ambipolar term. As we said, this numerical technique allows us to speed up, in a easy and efficient way, explicit time-stepping schemes for parabolic problems.

Besides, the Hall coefficient, $\eta_H$ dominates over the other two coefficients ($\eta$ and $\eta_A$) over most of the photosphere and then, the Hall effect can dominate the system
\citep{2014PhPl...21i2901K}. In this case, the system becomes skew-dominant, and the integration can not be done using the STS approach because it is unstable. In this situation, the system's eigenvalues become complex and to fulfil the CFL condition we would require that our time-step gets close to zero to maintain stability \citep{2006MNRAS.366.1329O, 2007MNRAS.376.1648O, Gurski:2011kf}. Thus, the proposed MHDSTS scheme uses yet another scheme, Hall Diffusion Scheme (HDS), which was proposed by \citet{2006MNRAS.366.1329O} to overcome this problem.

The STS technique has been used in different fields successfully, for example: \citet{2009ApJS..181..413C} described the implementation of the ambipolar term into a multidimensional MHD code based in a Total Variation Diminishing (TVD) scheme showing its use to follow complex MHD flows such as molecular cloud cores or protostellar discs; \citet{Grewenig:2010ex} introduced STS into a new class of numerical schemes called Fast Explicit Diffusion (FED) schemes to speed up the application of anisotropic diffusion filters for image enhancement or image compression tasks; \citet{2013MNRAS.434.2593T} presented an explicit scheme for Ohmic dissipation with smoothed particle magnetohydrodynamics (SPMHD) and used the STS technique to solve the Ohmic part of the induction equation; \citet{2013ApJ...763....6T} also adopted the technique to solve the Ohmic part of the induction equation in problems of protostellar collapse; \citet{2014MNRAS.444.1104W} provided an implementation of the STS technique into the \phantoms code \citep{2010MNRAS.406.1659P, 2010ASPC..429..274P, 2010MNRAS.405.1212L, Price:2017tl} for solving the ambipolar part of the induction equation. At this point we also mention that there is another family of STS scheme based on the Legendre polynomials and its properties which have been applied in problems involving thermal conduction in astrophysical plasmas as prominences or coronal rain successfully \citep{2012MNRAS.422.2102M, 2014JCoPh.257..594M, Xia:2016ca, 2017A&A...603A..42X}.

Equation (\ref{eq:conser}) can be split into different terms (ideal MHD, Ohmic diffusion, ambipolar diffusion, etc.), and these can be grouped together in a number of operators. There are a few options to group those terms into operators, but for our new  MHDSTS scheme we have grouped them in three operators (MHD, STS, and HDS) as described in the following sections. The splitting technique we have chosen to solve the system is the Strang splitting. This splitting is written in a way that keeps a second order accuracy along the simulation by permuting the order in which we apply the operators \citep{1968SJNA....5..506S, LeVeque:2002th}.

%%%%%%%%%%%%%%%%%%%%%%%%%%%%%%%%%%%%%%%%%%%%%%%%%%%%%%%%%
\subsubsection{MHD operator}\label{sec:mhd}
%%%%%%%%%%%%%%%%%%%%%%%%%%%%%%%%%%%%%%%%%%%%%%%%%%%%%%%%%

This operator is used to integrate the ideal MHD equations together with the battery term. We decided to include the non-ideal contribution of the Battery term into this solver because it does not limit the time step in solar-like problems \citep{2014PhPl...21i2901K}. This operator can be defined as:
%%%
\begin{equation}
    \dt{\mathbf{u}} = \mathcal{M}(\mathbf{u}) = L^{\mathrm{iMHD}}(\mathbf{u}) + L^{\mathrm{Batt}}(\mathbf{u}) \coma
\end{equation}
%%%
where the vector $\mathbf{u}$ is, as before, the set of conserved variables, $L^{\mathrm{iMHD}}(\mathbf{u}) = -\nabla \mathbf{F}(\mathbf{u}) + \mathbf{S}(\mathbf{u})$ is (as per equation (\ref{eq:conser})) the standard operator for ideal MHD, and $L^{\mathrm{Batt}}(\mathbf{u})$ is the operator for the battery term.

\bigskip

The temporal integration of this operator follows the same equations as (\ref{eq:rkm}), but in this case the operator being $\mathcal{M}(\mathbf{u})$ instead of $\mathcal{R}(\mathbf{u})$ and the time-step being $\Delta t_\mathrm{MHD}$, calculated as:
\begin{equation}\label{eq:dt_mhd}
    \Delta t_\mathrm{MHD}  = dt_\mathrm{iMHD} \coma%= \frac{(1/dx^2+1/dy^2+1/dz^2)^{-1/2}}{\max{(c_S,c_A)}}
\end{equation}
where $dt_\mathrm{iMHD}$ is obtained from the CFL condition applied to the operator $L^{\mathrm{iMHD}}(\mathbf{u})$. The Battery term time step, $dt_\mathrm{Batt} = \min(dx^2, dy^2, dz^2)/\eta_B$, doesn't play a roll into this operator due to its source term behaviour. 
It is important to point out that if there are no non-ideal terms this operator becomes a RK operator because there is no need to apply the Strang splitting. In this case, selecting $m=4$, we recover the RK scheme. This standard part of the \mancha code was throughly verified via either specific numerical tests, applied in 1D, 2D and 3D \citep{2010ApJ...719..357F}, or through the scientific numerical simulations by the code published in the literature in the last 10 years.

%%%%%%%%%%%%%%%%%%%%%%%%%%%%%%%%%%%%%%%%%%%%%%%%%%%%%%%%%
\subsubsection{STS operator}\label{sec:sts}
%%%%%%%%%%%%%%%%%%%%%%%%%%%%%%%%%%%%%%%%%%%%%%%%%%%%%%%%%

This operator is used to treat only the contribution of the Ohmic and Ambipolar terms and it can be written as:
%%%
\begin{equation}
    \dt{\mathbf{w}} = \mathcal{S}(\mathbf{w}) = L^{\mathrm{Ohmic}}(\mathbf{w}) + L^{\mathrm{Ambi}}(\mathbf{w}) \coma
\end{equation}
%%%
where the vector $\mathbf{w}$ corresponds to the conserved variables $[\mathbf{B},e](\mathbf{r},t)$ or $[\mathbf{B},e_\mathrm{int}](\mathbf{r},t)$ in the case where we prefer to use the internal energy equation, $L^{\mathrm{Ohmic}}(\mathbf{w})$ is the operator for Ohmic diffusion, and $L^{\mathrm{Ambi}}(\mathbf{w})$ is the operator for ambipolar diffusion.

The main idea behind STS is to demand stability not at each time step, but at the end of a bigger step called \emph{super-step} ($\Delta t_\textrm{STS}$). The $\Delta t_\textrm{STS}$ is calculated as a sum of \emph{sub-steps} $\tau_j$. To maximise the size and stability of the super-step, the small sub-steps are obtained using the optimality properties of modified Chebyshev polynomials \citep{Alexiades:1996vj}.
%%%
\begin{equation}\label{eq:tauj}
    \tau_j=dt_{\textrm{dif}}\left[(\nu-1)\cos\left(\frac{2j-1}{N_\textrm{STS}}\frac{\pi}{2}\right)+1+\nu\right]^{-1} \coma
\end{equation}
%%%
in this expression, $dt_\textrm{dif}$ is the minimum time step given by the Ohmic term ($dt_\textrm{Ohmic} = \min(dx^2, dy^2, dz^2)/\eta$) and by the ambipolar term ($dt_\textrm{Ambi} = \min(dx^2, dy^2, dz^2)/\eta_A$). $N_\textrm{STS}$ is the number of sub-steps, and $\nu$ is a damping parameter, associated to the Chebyshev polynomials, to reduce higher frequencies and obtain a strong stability condition for the numerical method. This parameter is within the interval (0,$\lambda_\textrm{min}/\lambda_\textrm{max}$], where $\lambda_\textrm{min}$ and $\lambda_\textrm{max}$ are the smallest and the bigger eigenvalues of the matrix $\mathcal{S}(\mathbf{w})$ respectively \citep{Alexiades:1996vj, Gurski:2011kf}. When $\nu$ takes values close to one, the system is highly damped, and the super-step becomes more accurate in the same way that any numerical explicit scheme has smaller errors resulting from small time steps. When $\nu$ takes values close to zero, the system becomes unstable.
 
Thus, the value of the parameters $(N_\mathrm{STS} ,\nu)$ control the stability, accuracy and speed of the method and depend on the spectral properties of the matrix $\mathcal{S}(\mathbf{w})$. But, as pointed out by \citet{Alexiades:1996vj}, a precise knowledge of the eigenvalues is not needed to obtain a robust method, so the values can be arbitrarily chosen by the user considering certain limits (stability, speed, error, etc.).\\
The time update by the STS operator can be written as:
%%%
\begin{equation}\label{eq:sts}
    \textbf{w}(\mathbf{r},t_{n+1}) = \textbf{w}(\mathbf{r},t_n)+\sum_{j=1}^{N_\textrm{STS}} \tau_j\frac{\partial\textbf{w}(\mathbf{r},t)}{\partial t}\Big|_{t_n+\sum_{k=1}^{j-1} \tau_k} \coma
\end{equation}
%%%
where $t_{n+1}= t_n+\Delta t_\textrm{STS}$ and the super-step $\Delta t_\textrm{STS}$ is given by
%%%
\begin{equation}\label{eq:superstep}
    \Delta t_\textrm{STS} = \sum_{j=1}^{N_\textrm{STS}} \tau_j=dt_\textrm{dif}F(N_\textrm{STS},\nu) \coma
\end{equation}
with
\begin{equation}
    F(N_\textrm{STS},\nu)=\frac{N_\textrm{STS}}{2\nu^{1/2}}\frac{\left(1+\nu^{1/2}\right)^{2N_\textrm{STS}} - \left(1-\nu^{1/2}\right)^{2N_\textrm{STS}}}{\left(1+\nu^{1/2}\right)^{2N_\textrm{STS}} + \left(1-\nu^{1/2}\right)^{2N_\textrm{STS}}} \punto
\end{equation}
%%%
In the limit when $\nu$ tends to zero, the  super-step formed by $N$ sub-steps covers an $N$ times bigger interval compared to $N$ explicit time steps, namely:
%%%
\begin{equation}\label{eq:flim}
    \lim_{\nu \to 0}F(N_\textrm{STS},\nu)=N_\textrm{STS}^2 \implies \lim_{\nu \to 0}\Delta t_\textrm{STS} = N_\textrm{STS}^2 dt_\textrm{dif} \punto
\end{equation}
%%%
The STS scheme is a first order scheme and it belongs to the family of Runge-Kutta-Chebyshev (RKC) methods with $N$ steps. Therefore, if used as described above, the overall precision of the MHDSTS scheme would be of first order, even if the precision of the other operators was bigger. In order to increase its order of accuracy one could perform a Richardson extrapolation \citep{Richardson:1911vw, 1927RSPTA.226..299R} as suggested by \citet{2006MNRAS.366.1329O}. However, this implies many executions of the STS operator: three executions to reach second order, seven executions to reach third order and fifteen executions to reach fourth order.

\noindent For that reason, instead of performing a Richardson extrapolation, we consider the STS scheme as an "Eulerian" first-order step of our multi-step RK scheme, so now we should reach fourth order just with four calls to the STS scheme. In other words, we are using the STS scheme to obtain the solution at the stable points of the RK scheme and in this way, get a high order solution. We will refer to this technique for increasing the order of a numerical scheme as \emph{RK-wrapper}. Thus, the complete temporal scheme for the STS operator can be written similarly to the equations (\ref{eq:rkm}) but, in this case, using  the operator $\mathcal{S}(\mathbf{w})$ and its corresponding time step $\Delta t_\textrm{STS}$: 
%%%
\begin{subequations}
    \begin{align}
        \mathbf{w}^{(k)} &= \mathbf{w}^{(n)}+\alpha_k\Delta t_\textrm{STS} \mathcal{S}\left(\mathbf{w}^{(k-1)}\right)\coma \quad k=1,\dots,m\coma \\
        \mathbf{w}^{(n+1)} &=  \mathbf{w}^{(m)}\punto 
    \end{align}
\end{subequations}
%%%
Each step of this RK-wrapper has a size of $\alpha_k\Delta t_\textrm{STS}$, with $\alpha_k$ given by eq. (\ref{eq:ak}). Each of these steps represents a whole STS super-step, where $dt_\textrm{dif}$ is scaled with $\alpha_k$ and the corresponding $\tau_j$ values are calculated according to equation (\ref{eq:tauj}). With these, the conserved variables of the vector $\mathbf{w}^{(k-1)}$ are calculated using the expression (\ref{eq:sts}).

%%%%%%%%%%%%%%%%%%%%%%%%%%%%%%%%%%%%%%%%%%%%%%%%%%%%%%%%%
\subsubsection{HDS operator}\label{sec:hds}
%%%%%%%%%%%%%%%%%%%%%%%%%%%%%%%%%%%%%%%%%%%%%%%%%%%%%%%%%
The HDS operator only solves the Hall term in the induction equation. It is designed by \citet{2006MNRAS.366.1329O} to overcome the problems originated by a skew-symmetric Hall term dominated system, and it can be written in conservative form as:
%%%
\begin{equation}
    \dt{\mathbf{b}} =\mathcal{H}(\mathbf{b}) = L^{\mathrm{Hall}}(\mathbf{b}) \coma
\end{equation}
%%%
where the vector $\mathbf{b}$ are the conserved variables $\mathbf{B}(\mathbf{r},t)$ and $L^{\mathrm{Hall}}(\mathbf{b})$ is the operator for the Hall term. This operator works similarly to the MHD operator described above. However, unlike it, the update of the magnetic field components is done using all the information available at the moment. \citet{2006MNRAS.366.1329O} applied the HDS scheme by using a first order RK scheme and then, increasing the accuracy order by the application of a Richardson extrapolation. Instead, we follow the same idea proposed in Section \ref{sec:sts}, treating the individual HDS operator as an "Eulerian'' step to use the RK-wrapper to increase its accuracy order. By doing so, we keep the stability properties of this implicit-like scheme but reaching a higher accuracy order in time and with less computational effort. The proposed scheme would be:
\begin{subequations}\label{eq:hds}
    \begin{alignat}{1}
%\mathbf{B}^{(0)} = {} & \mathbf{B}^{(n)}\coma \\
        B^{(k)}_{x} = {} & B^{(n)}_{x}+\alpha_k dt_\textrm{Hall}\mathcal{H}\left(B^{(k-1)}_{x},B^{(k-1)}_{y},B^{(k-1)}_{z}\right)\coma \\
        B^{(k)}_{y} = {} & B^{(n)}_{y}+\alpha_k dt_\textrm{Hall}\mathcal{H}\left(B^{(k)}_{x},B^{(k-1)}_{y},B^{(k-1)}_{z}\right)\coma \\
        B^{(k)}_{z} = {} & B^{(n)}_{z}+\alpha_k dt_\textrm{Hall}\mathcal{H}\left(B^{(k)}_{x},B^{(k)}_{y},B^{(k-1)}_{z}\right)\coma      \scalebox{0.7}{$~k=1,\dots,m$}\coma \\
        \mathbf{B}^{(n+1)} = {} & \mathbf{B}^{(m)} \punto 
    \end{alignat}
\end{subequations}

In these equations $dt_\textrm{Hall}$ is the time step imposed by the Hall term through the CFL condition given by \citet{2007MNRAS.376.1648O}, that is, $dt_\mathrm{Hall} = 2/\sqrt{27}\min(dx^2, dy^2, dz^2)/\eta_H$. The HDS operator advances in time $\Delta t_\textrm{HDS} = N_\textrm{HDS} dt_\textrm{Hall}$, by repeating the execution of the equations (\ref{eq:hds}) $N_\textrm{HDS}$ times, which is the number of sub-cycles needed to cover one super-step $\Delta t_\textrm{STS}$ and/or one $\Delta t_\textrm{MHD}$; see next section for details.

If the order in which the equations (\ref{eq:hds}a, \ref{eq:hds}b, and \ref{eq:hds}c) are solved is kept for successive time steps, an artificial handedness is introduced into the scheme. This could be avoid by reversing the order between time step or, as it was pointed by \citet{2007MNRAS.376.1648O}, performing a random permutation of the order in which the magnetic field components (\ref{eq:hds}a, \ref{eq:hds}b, and \ref{eq:hds}c) are solved over successive time steps. This is especially important under certain circumstances, as for example if a strong directional bias is introduced in the initial state or in 3D numerical experiments. Such permutation can result in a small loss of numerical stability.

%%%%%%%%%%%%%%%%%%%%%%%%%%%%%%%%%%%%%%%%%%%%%%%%%%%%%%%%%
\subsection{All together: timing and accuracy order}
%%%%%%%%%%%%%%%%%%%%%%%%%%%%%%%%%%%%%%%%%%%%%%%%%%%%%%%%%
This new scheme is only worthwhile with respect to the RK when the Ambipolar or Hall terms introduce a heavy constraint in the time step as we commented in Sections \ref{sec:sts} and \ref{sec:hds}. Then, one should select the time steps for the operators according to the following rule:
\begin{equation}\label{eq:timing}
    dt_\textrm{dif} \le \Delta t_\textrm{STS}(N_\textrm{STS},\nu) = \Delta t_\textrm{HDS}(N_\textrm{HDS}) \le \Delta t_\textrm{MHD} \punto
\end{equation}
\begin{figure}[t]
    \resizebox{\hsize}{!}{\includegraphics{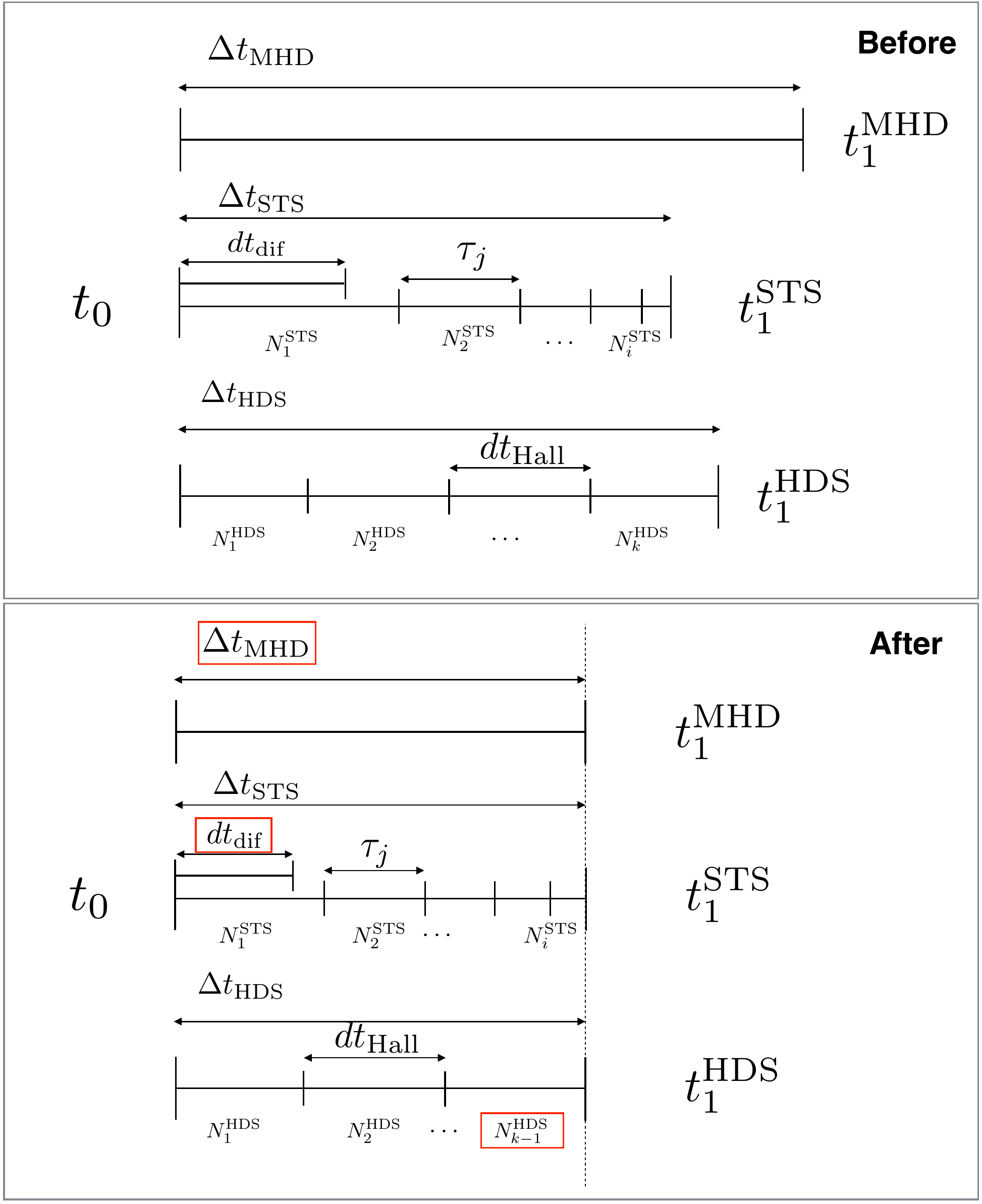}}
    \caption{\footnotesize Example of time step adjusting. In the upper panel it is shown the situation before the adjust, all the super steps are different. In the lower panel the red square shows in each case the quantity that was changed. In this case, the code starts removing one sub-cycle of the HDS, then reduces $dt_\mathrm{dif}$ to mach the HDS super-step and finally reduces the MHD super-step to mach the other two.}
    \label{fig:timing}
\end{figure}
\begin{figure}[t]
    \resizebox{\columnwidth}{!}{\includegraphics{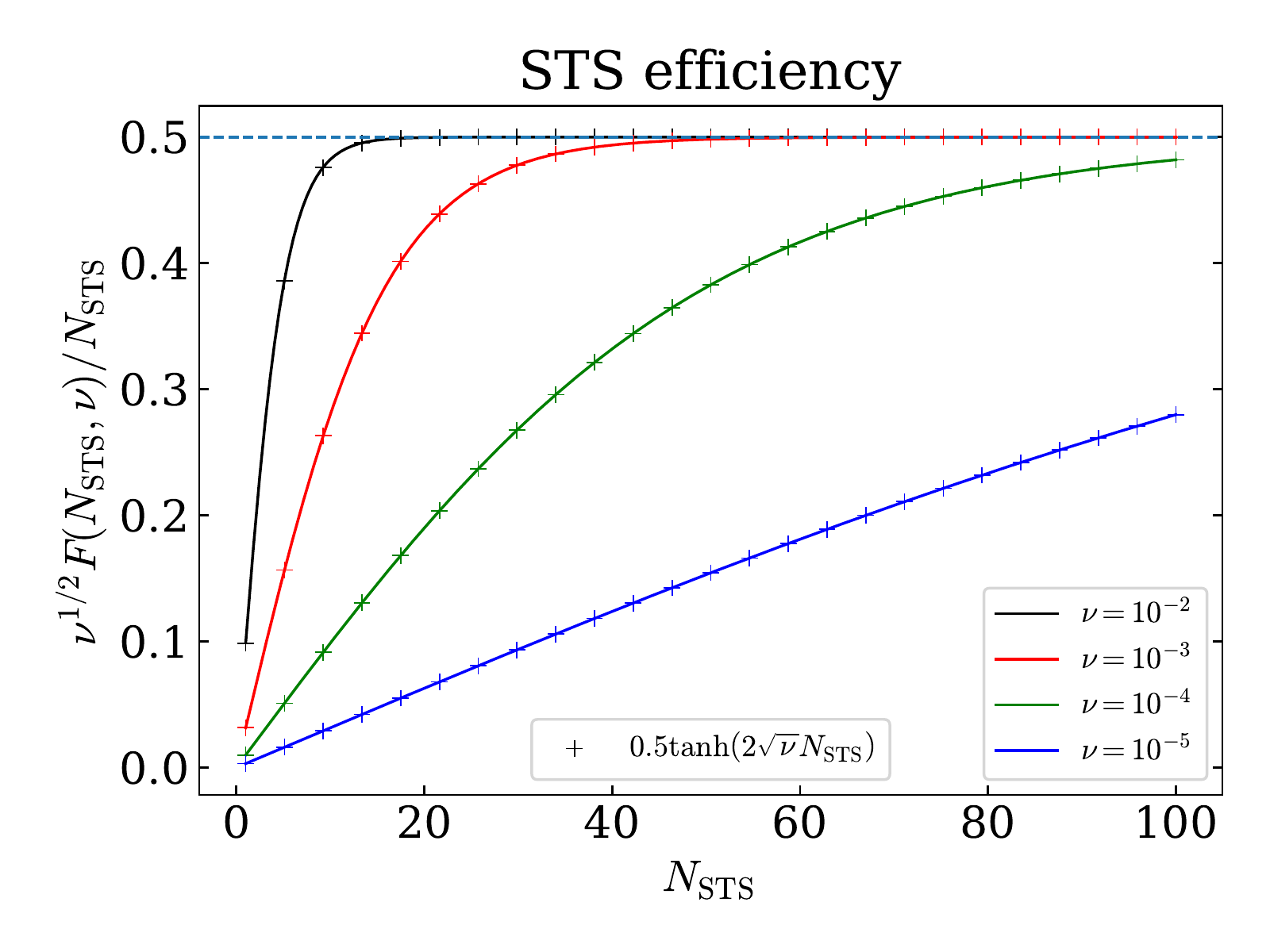}}
    \caption{\footnotesize STS efficiency ($E_\textrm{STS}$) defined by equation (\ref{eq:eff}) for different values of $\nu$ and its approximation ("+" markers). The dashed line indicates the asymptotic value of $E_\textrm{STS}$ when $N_\textrm{STS} \to \infty$.}
    \label{fig:eff}
\end{figure}
When the requirements of  $dt_\textrm{dif} \le \Delta t_\textrm{MHD}$ is not fulfilled, the evolution of the system is dominated by hyperbolic ideal MHD terms, and there is no need to apply this new MHDSTS scheme, then we are out of the STS (and/or HDS) regime.

To choose the time step for each iteration, one first has to choose the pair ($N_\textrm{STS},\nu)$ to keep the accuracy and stability during the simulation and also to try to maximise the size of the super-step along the simulation.\\
One option for choosing those values can be obtained using equation (\ref{eq:flim}) and forcing the STS operator to jump in time as far as the MHD operator allows it, so in the limit when $\nu \to 0$, we can write that
\begin{equation}\label{eq:optimun_n}
    \Delta t_\textrm{STS} \approx dt_\textrm{dif} N_\textrm{STS}^2 = \Delta t_\textrm{MHD}  \implies  N_\textrm{STS} %\approx \textrm{Floor}\left(\sqrt{\frac{\Delta t_\textrm{MHD}}{dt_\textrm{dif}}} \right)\punto
    \approx \floor*{\sqrt{\frac{\Delta t_\textrm{MHD}}{dt_\textrm{dif}}}} \coma
\end{equation}
getting in this way an optimum approximation for $N_\textrm{STS}$ which should be very close to the number of sub-steps needed to cover the time step imposed by the ideal MHD CFL.\\
Selecting the parameter $\nu$ can be a bit tricky, and for that reason it is convenient to define the STS efficiency as:
\begin{equation}\label{eq:eff}
    E_\textrm{STS}=\nu^{1/2}\frac{\Delta t_\textrm{STS}}{dt_\textrm{dif} N_\textrm{STS}}=\nu^{1/2}\frac{F(N_\textrm{STS},\nu)}{N_\textrm{STS}} \punto
\end{equation}
This expression reaches an asymptotic value of $0.5$ when $N_\textrm{STS} \to \infty$ for a given $\nu$, see dashed line in Fig.~\ref{fig:eff}. It can be also approximated as $0.5\tanh{\left(2\sqrt{\nu}N_\textrm{STS}\right)}$, see "+" markers in Fig.~\ref{fig:eff}. Using this approximation for the efficiency we can obtain a value of $\nu$ to achieve a high efficiency and to maintain the stability and accuracy of the STS scheme. To do so, we can impose a certain efficiency and calculate, for a chosen $N_\textrm{STS}$ what value of $\nu$ it corresponds to. After several tests we have concluded that a good compromise between speed and stability is reached fixing the STS efficiency around a 76\% of its maximum. This is equivalent to equating $\tanh$ to one, in which case we can write $\nu \approx 0.25/N_\textrm{STS}^2$. Finally, replacing the value of $N_\textrm{STS}$ obtained from eq. (\ref{eq:optimun_n}) we get an estimation for $\nu$. 
The choice of number of HDS sub-cycles $N_\textrm{HDS}$ is done in execution time as $N_\textrm{HDS} = \floor*{\Delta t_\textrm{STS}/dt_\textrm{Hall}}$.

The upper limit in time for the integration time step at any moment of simulation is given by the CFL condition applied to the ideal MHD operator $\Delta t_\textrm{MHD}$. 
The goal then is to adjust $dt_\mathrm{iMHD}$, $dt_\textrm{dif}$ and $N_\textrm{HDS}$ to fulfil the identity $\Delta t_\textrm{MHD} = \Delta t_\textrm{STS} = \Delta t_\textrm{HDS}$ (see Fig. \ref{fig:timing} for an example).

Eventually, as we have mentioned, due to the Strang splitting, the overall order of the MHDSTS scheme is limited to second order whenever each operator reaches second order accuracy  or higher using the RK-wrapper method. This accuracy is expected to decrease due to the different combinations of values we can choose for the pair $(N_\textrm{STS},\nu)$ and the physical characteristics of each simulation.

%%%%%%%%%%%%%%%%%%%%%%%%%%%%%%%%%%%%%%%%%%%%%%%%%%%%%%%%
\section{Numerical tests}\label{sec:test}
%%%%%%%%%%%%%%%%%%%%%%%%%%%%%%%%%%%%%%%%%%%%%%%%%%%%%%%%%

As mentioned in Section \ref{sec:mhd}, when all the non-ideal terms are switched off the MHDSTS scheme becomes a pure RK scheme because there is no need to apply the Strang splitting. This means that all the previous results obtained with \mancha as well as the tests presented in \citet{2010ApJ...719..357F} are also valid for this new version.

Bearing this in mind, the sections below present specific numerical tests to only validate the newly introduced operators either individually, or working all together.

%%%%%%%%%%%%%%%%%%%%%%%%%%%
\subsection{MHD operator test}\label{sec:MHD_op}
%%%%%%%%%%%%%%%%%%%%%%%%%%%
For testing the MHD operator applied to the ideal part of the equations and the battery term, we performed the numerical experiment proposed by \citet{2012JCoPh.231..870T}. In this test, a magnetic field is generated from scratch by means of the Biermann battery term in the induction equation. For this test, we have removed all the artificial diffusivities and considered  only the contribution of the battery term.
\begin{figure}[t]
    \resizebox{\hsize}{!}{\includegraphics{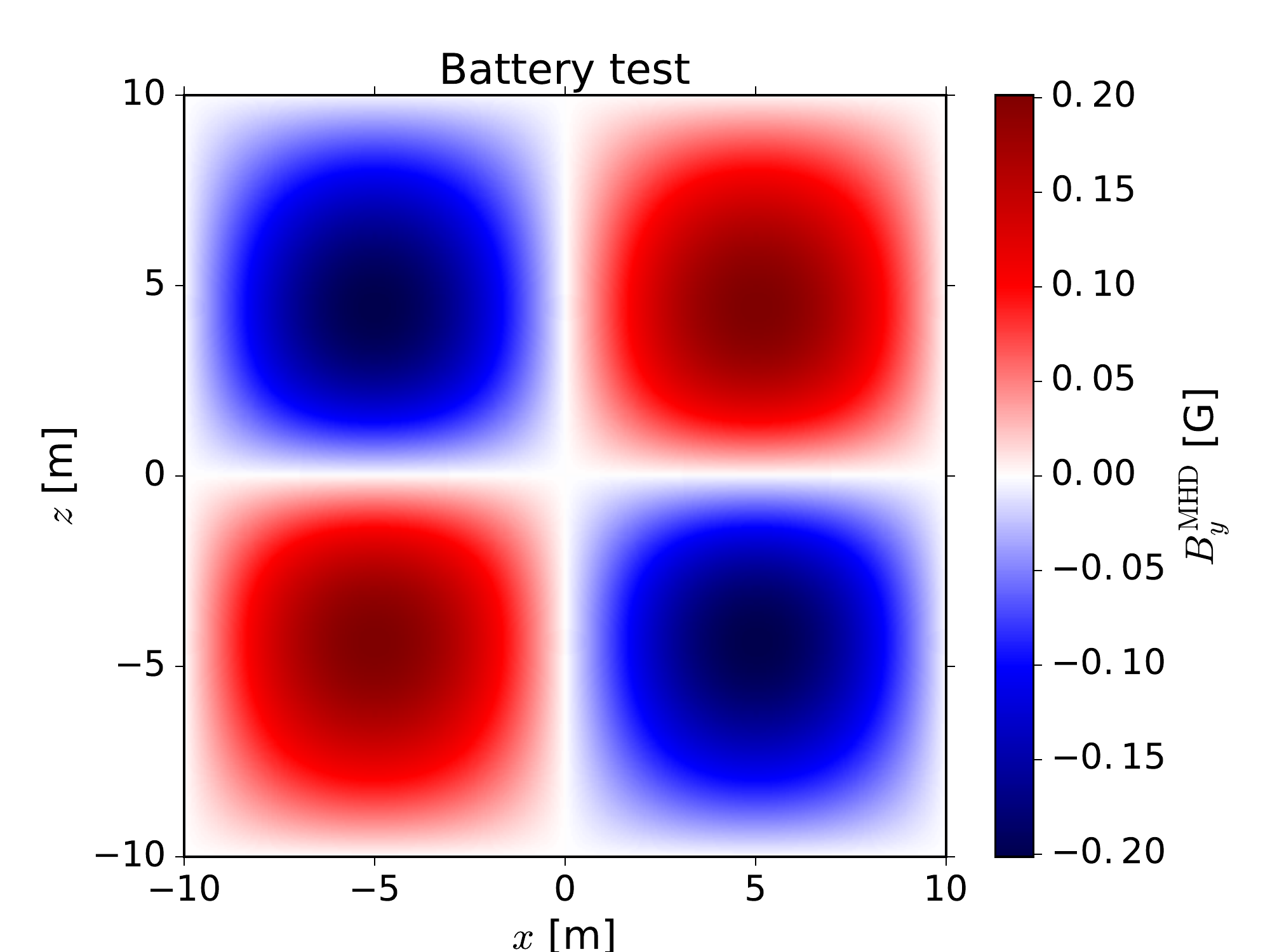}}
    \caption{\footnotesize Biermann battery test. The image shows the $y$ component of the magnetic field vector at one time step ($dt= 0.02$ s) after the beginning of the simulation.}
    \label{fig:bat}
\end{figure}

\begin{figure}
    \centering
    \includegraphics[width=1\columnwidth]{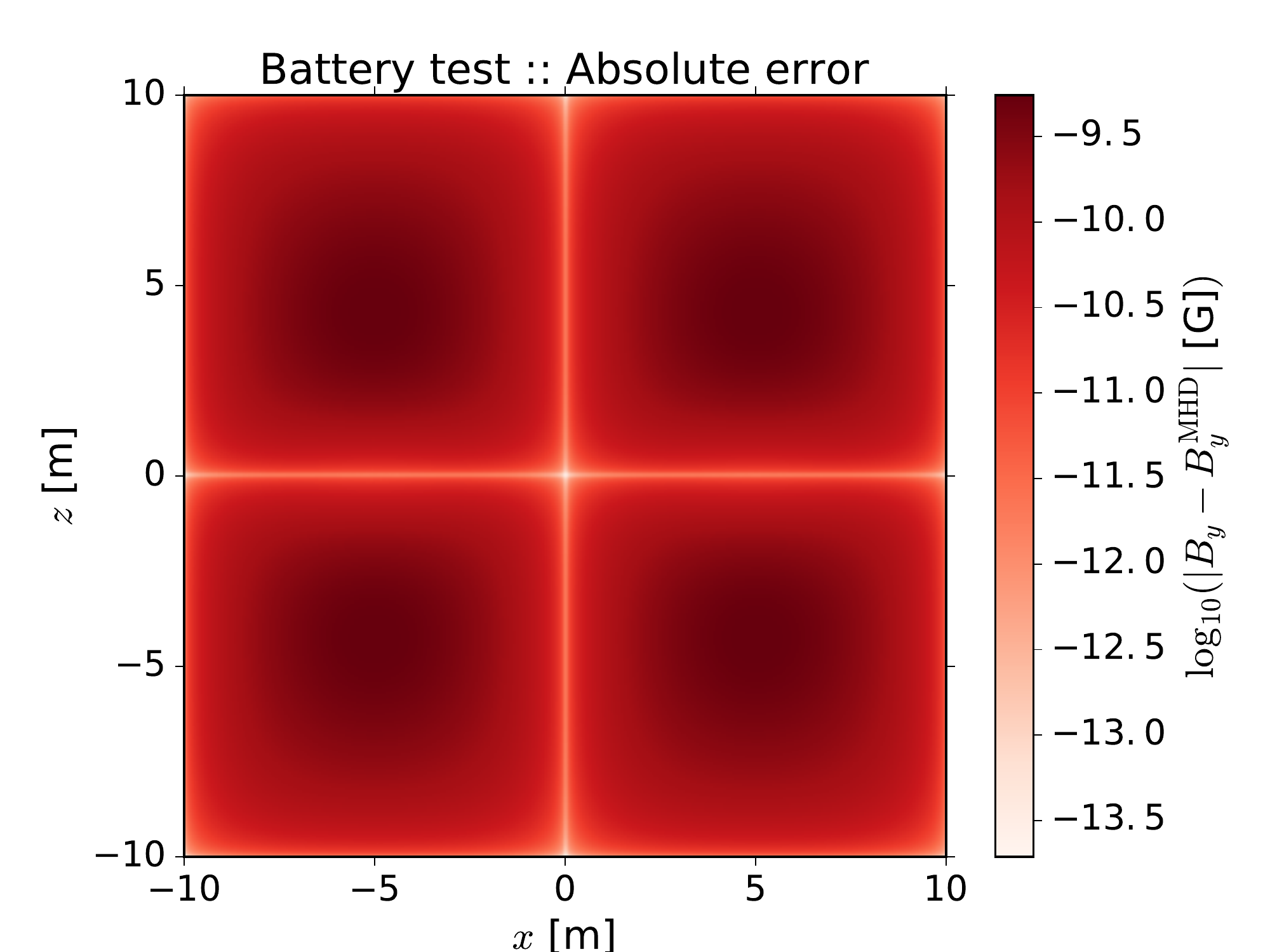}
    \caption{\footnotesize Logarithm of the absolute error in the $B_y$ component of the magnetic field between the analytical solution and the numerical solution given by the MHD operator. The format of the figure is the same as Fig. \ref{fig:bat}. The error is minimum at locations with maximum electron density $n_e$.}
    \label{fig:bat_err}
\end{figure}

The initial conditions for this test consist in a homogeneous isothermal background without magnetic field and no velocity fields ($\mathbf{B}=\mathbf{v}=0$). To initiate the evolution, a small perturbations in electron density (\ref{eq:ne}) and electron pressure (\ref{eq:pe}) are applied.
\begin{equation}\label{eq:ne}
    n_e=n_0+n_1\cos(k_xx) \coma
\end{equation}
\begin{equation}\label{eq:pe}
    p_e=p_0+p_1\cos(k_zz) \punto
\end{equation}
The analytical solution (\ref{eq:bat_by}) is obtained by introducing the initial condition into the MHD equations and integrating it one single time step.
\begin{equation}\label{eq:bat_by}
    \frac{\partial B_y}{\partial t} = \frac{n_1p_1k_xk_z\sin(k_xx)\sin(k_zz)}{e[n_0+n_1\cos(k_xx)]^2} \punto
\end{equation}
The numerical set-up consists in a 2.5D simulation box in the plane $x-z$. For this experiment we considered a double periodic domain in both $x$ and $z$ coordinates, covering a physical space of $\pm$ 10 meters, and using the constants, in S.I, $n_0=1/e$, $p_0=1$, $n_1=0.1/e$, $p_1=0.1$, $k_x=k_z=2\pi/L$, and $L=20$.

The numerical solution is obtained using our MHD operator, Fig. \ref{fig:bat}, and it is compared with the analytical solution by calculating the absolute error between them, see Fig. \ref{fig:bat_err}. We only compare the numerical solution with the analytical one after one single time step because if the system keeps evolving the analytical solution (\ref{eq:bat_by}) is not valid anymore, in other words, the convective term will contribute to the solution. It can be seen that most of the numerical domain does not exceed $10^{-9.3}$ G of absolute error and this is concentrated in the regions where the initial electron density and electron pressure perturbations are maximum.

%%%%%%%%%%%%%%%%%%%%%%%%%%%%%%%%%%%%%%%%%%%%%%%%%%%%%%%%%
\subsection{STS operator test}\label{sec:STS_op}
%%%%%%%%%%%%%%%%%%%%%%%%%%%%%%%%%%%%%%%%%%%%%%%%%%%%%%%%%

In order to test the implementation of the STS operator we use an experiment of Alfv\'en wave decay under the sole action of the ambipolar diffusion as it was proposed by \citet{2009ApJS..181..413C}. This experiment is based on the analytical derivation of a dispersion relation characterising the propagation of an Alfv\'en wave in a homogeneous partially ionised plasma permeated by a horizontal and homogeneous magnetic field done by \cite{1996ApJ...465..775B} when a strong coupling approximation is taken into account.
\begin{equation}\label{eq:disp}
    \omega^2+ik^2\eta_A\omega-k^2c_A^2=0 \coma \\
\end{equation}
In this relation, $c_A=B/\sqrt{\rho \mu_0}$ is the Alfv\'en speed, $k$ is a real wavenumber and $\omega = \omega_R+i\omega_I$ is the complex angular frequency. It is known that the action of the ambipolar diffusion adds an imaginary term into the dispersion relation, which physically means that the Alfv\'en wave is unable to propagate when $k\ge 2c_A/\eta_A$ \citep{1969ApJ...156..445K}. Under this strong coupling approximation and considering an ideal isothermal MHD system we can follow the evolution in time of a standing wave and compare the numerical results with the analytical solution of its first normal mode \citep{1970AmJPh..38..666M}
\begin{equation}\label{eq:1mode}
    h(t) = h_0|\sin(\omega_R t)|\exp(\omega_I t) \coma
\end{equation}
where $h_0$ is the initial wave amplitude.

\bigskip

For our experiments, we have considered an isothermal system in a 2.5D periodic domain in the plane $x-z$ of side $L$ with a constant ambipolar coefficient and a constant magnetic field $B_0$ along the $x$ coordinate, introducing a standing wave along the $z$ coordinate with an initial velocity
\begin{equation}\label{eq:ini_cond_sts}
    \vec{\mathbf v}=\mathrm{v}_\mathrm{amp} c_A \sin(k_x x)\hat{\mathbf k} \coma
\end{equation}
where $\mathrm{v}_\mathrm{amp}$ is a dimensionless initial peak amplitude. Taking $k=k_x$ in (\ref{eq:disp}), it can be obtained that
\begin{equation}
    \omega_R = \frac{k_x}{2}\sqrt{4c_A^2-k_x^2\eta_A^2}, \quad \omega_I=-\frac{k_x^2\eta_A}{2} \punto
\end{equation}
The analytical solution can be written as a superposition of two waves moving in opposite directions:
\begin{equation}\label{eq:ambi_sol}
    B_{z}(x,t)=A_0\cos{(k_x x)}\sin{(\omega_R t)} \exp{(\omega_I t)}\coma
\end{equation}
with
\begin{equation}
    A_0 = \frac{\mathrm{v}_\mathrm{amp}c_A k_x B_0}{\omega_R} \coma
\end{equation}
and tracking the evolution of the Root-Mean-Square (RMS) magnetic field $<B_{z}^2(t)>^{1/2}$, the initial amplitude in (\ref{eq:1mode}) is $h_0 = A_0/\sqrt{2}$.

%%%%%%%%%%%%%%%%%%%%%%%%%%%%%%%%%%

We set the parameters  $\mathrm{v}_\mathrm{amp}=0.01$, $L=1$ m, $B_0=\mu_0^{1/2}$ T, $p=1$ N/m, $\rho=1$ kg/m$^3$, and $T=1$ K for the initial condition and background. We considered four cases with different constant ambipolar coefficients $\eta_A=0.01, 0.03, 0.1,$ and $0.3$ m$^2$ s$^{-1}$. 

\begin{figure*}
    \centering
	\includegraphics[width=0.49\textwidth]{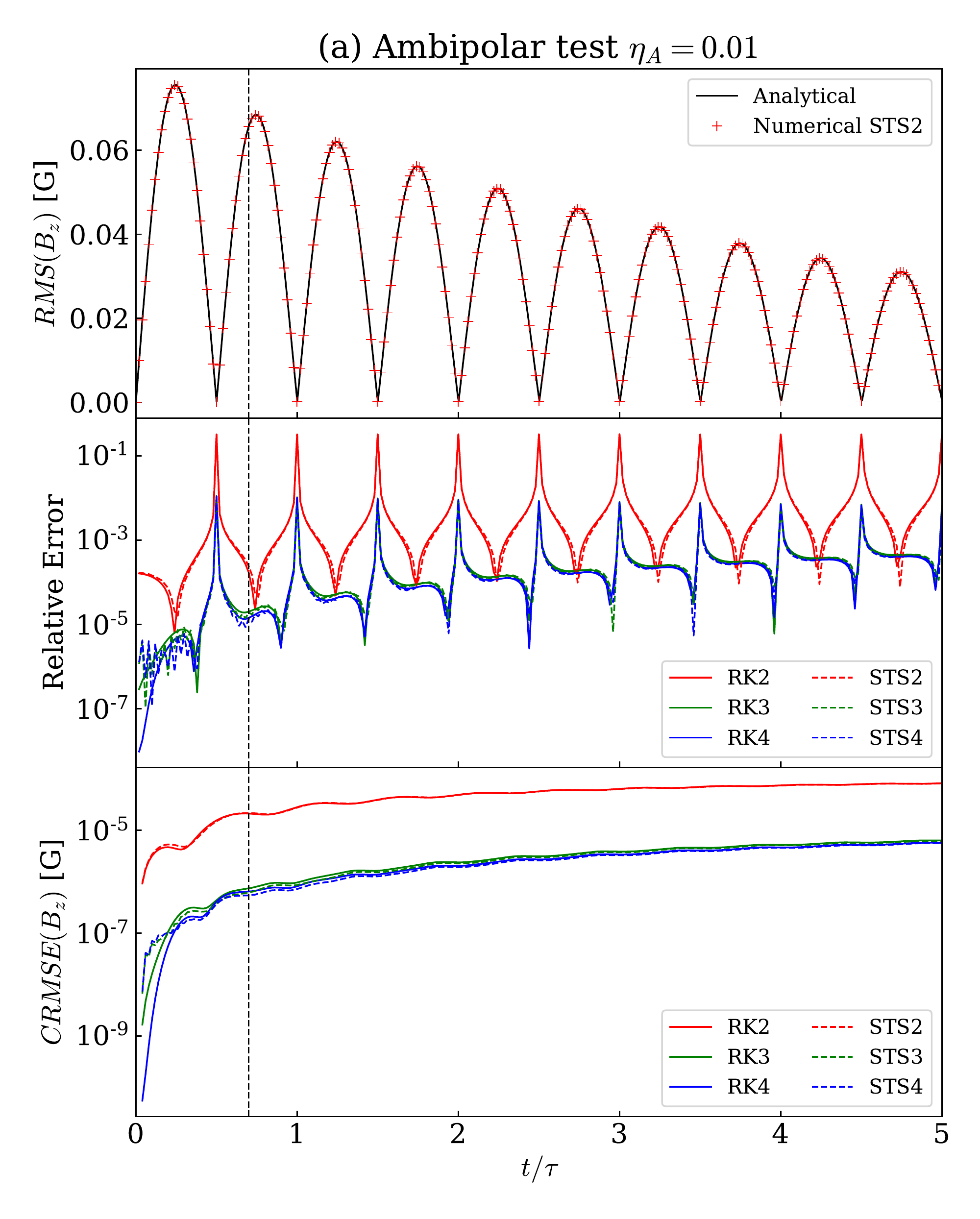}
	\includegraphics[width=0.49\textwidth]{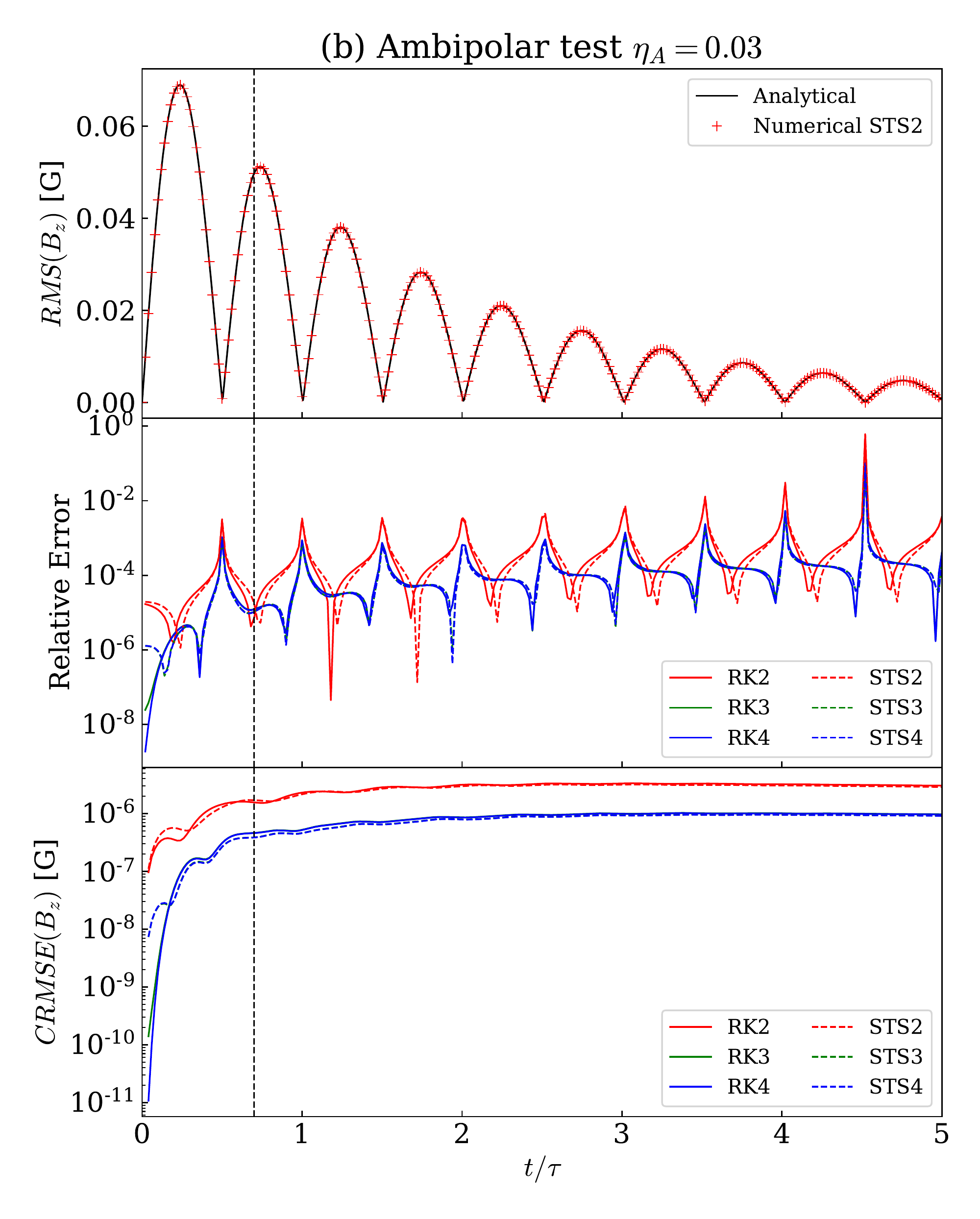}\\
	\vspace{-0.3cm}
	\includegraphics[width=0.49\textwidth]{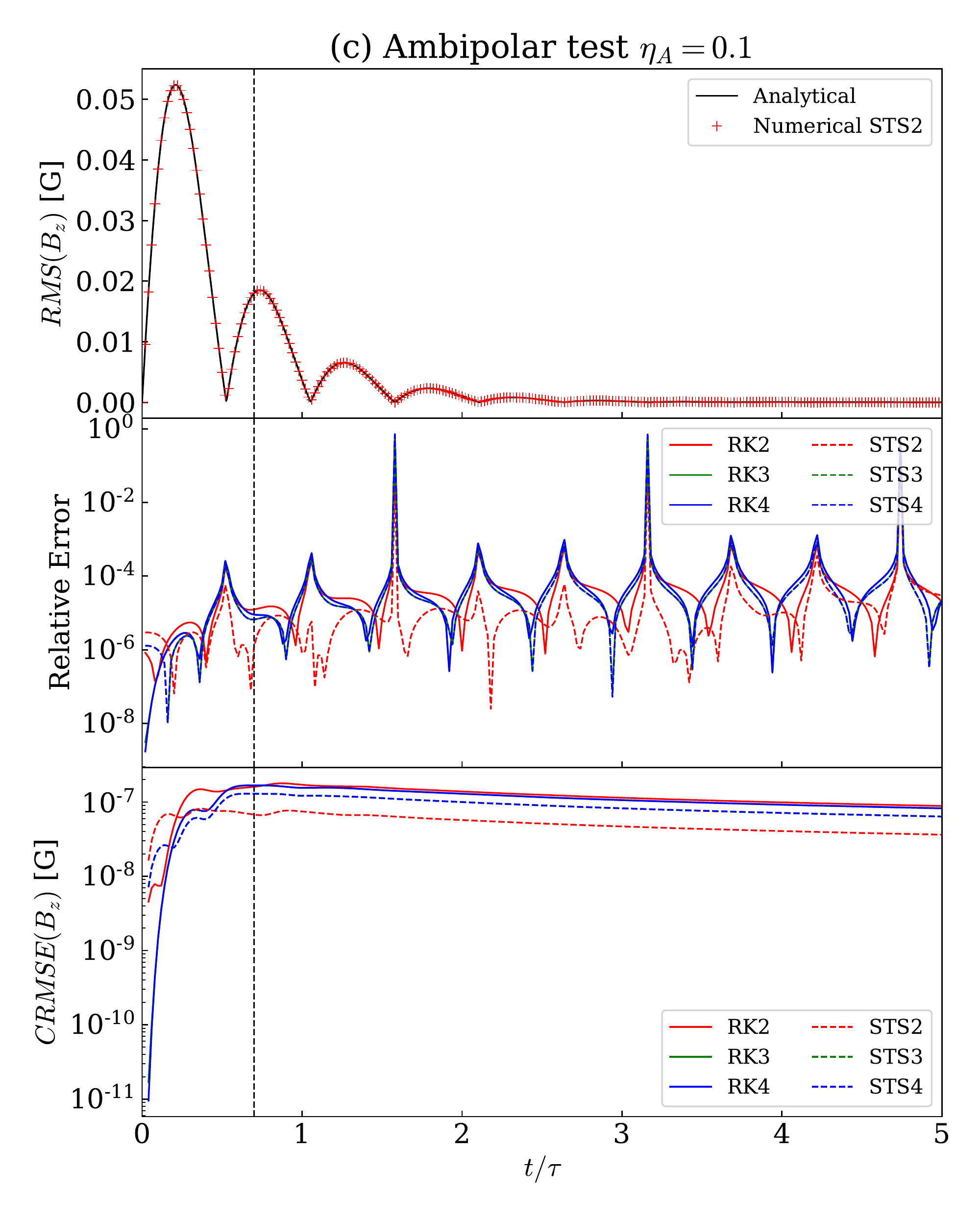}
	\includegraphics[width=0.49\textwidth]{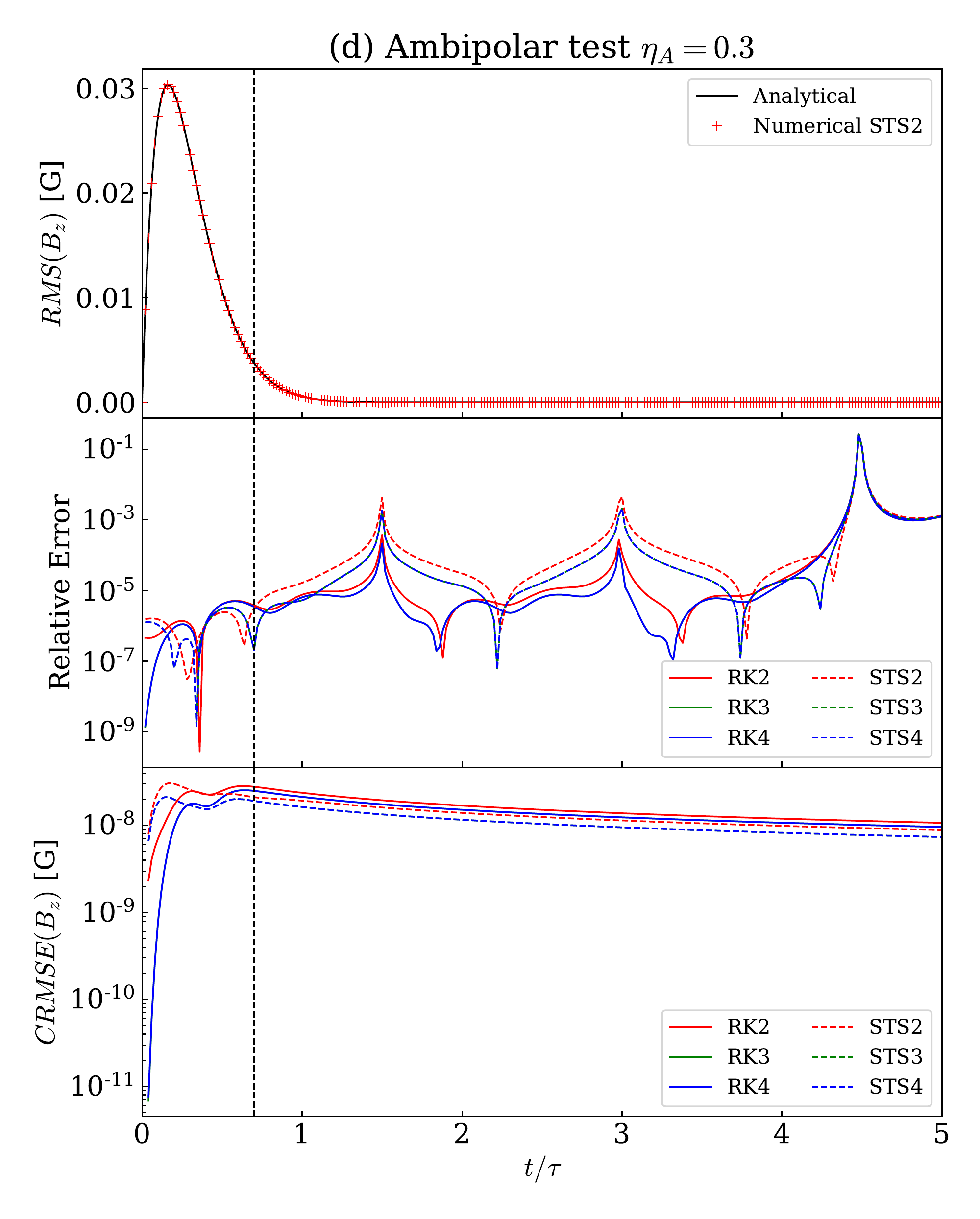}
	\caption{\footnotesize The upper panel shows the RMS value of the $B_z$ component of the magnetic field produced by the Alfv\'en wave, as a function of time. The solid black line corresponds to the analytical solution and the crosses are the numerical solution obtained with the STS scheme at order 2. It can be seen there is good agreement with the analytical solution for all the diffusivities tested. The middle panel shows the relative error in time for the RMS value of $B_z$ component. The red, green and blue lines are for the numerical solution given by the RK scheme working in the second, third and fourth order, respectively. The dashed lines corresponds to the solution obtained with the STS operator working also in second, third and fourth order. The lower panel shows the CRMSE in time for both schemes defined by eq. (\ref{eq:ce}). As we expect, the behaviour of both schemes is similar as it is shown by the CRMSE. The time axis is in the units of the crossing time $\tau=L/c_A$ and the dotted vertical lines mark the time where the $L_1$-norm is calculated, $t_0=0.7\tau$.}
	\label{fig:ambitest}
	\vspace{-0.5cm}
\end{figure*}

First, we used $N_\textrm{STS}=1$ and $\nu=0$ for the STS scheme to test how the error changes by increasing the order when using the operator splitting. In Figure \ref{fig:ambitest} we can see the results obtained in each case. In the upper panel of each subfigure we have the RMS evolution in time for the $B_z$ component. In all cases the agreement between the analytical and numerical solutions is good, notice we only plot the numerical solution when the order is set to 2. In the middle panel the relative error in time for the numerical solution obtained with the RK scheme and the STS scheme and three different orders is plotted. If we focus in the solution at order 2, we can see how the shape and values of each schemes are very similar in time. The same can be seen for the other orders.
Because the relative error goes to infinity when $<B^2_z(t)>^{1/2}$ drops to zero a good tool to see the time evolution of the error can be the Cumulative Root-Mean-Square Error (CRMSE) in time define as
\begin{equation}\label{eq:ce}
    \mathrm{CRMSE}(t)=\sqrt{\frac{1}{n}\sum_{i=1}^{n}\left[ f(t_i)_\mathrm{Analytical} -f(t_i)_\mathrm{Numerical} \right]^2}\coma
\end{equation}
where $n$ is the number of snapshots saved up to time $t_n$, $t_i$ is the time of each output, and $f(t_i) = <B^2_z(x,t_i)>^{1/2}$ \citep{2014MNRAS.444.1104W}. This quantity represents the sample standard deviation of the differences between the RMS value from our numerical experiment and the analytical solution and, as \citet{Hyndman:2006dt} point, it is a good measure of the accuracy between models at the same scale. This function is plotted in the lower panel of the subfigures of the Figure \ref{fig:ambitest}.
When the rate at which the scheme adds error to the RMS is slower than the growth of the factor $\sqrt{n}$ for increasing time, the CRMSE will decrease slowly with time to an asymptotic value. This is clearly seen when the wave is damped quickly due to the high ambipolar diffusion. Furthermore, it is clearly seen how both schemes have a similar behaviour for all the ambipolar diffusivities tested.

Now we study how the pair $(N_\mathrm{STS},\nu)$ affects to the error, computational time and accuracy order of the STS scheme in comparison with the RK. To do this, we made a couple of simulations using different values of them. We took three values of $\nu$ ($0.1, 0.01$ and $0.001$) and for each of them, $N_\mathrm{STS}$ was taken between one and six.

\begin{figure*}
	\centering
    \includegraphics[width=0.49\textwidth]{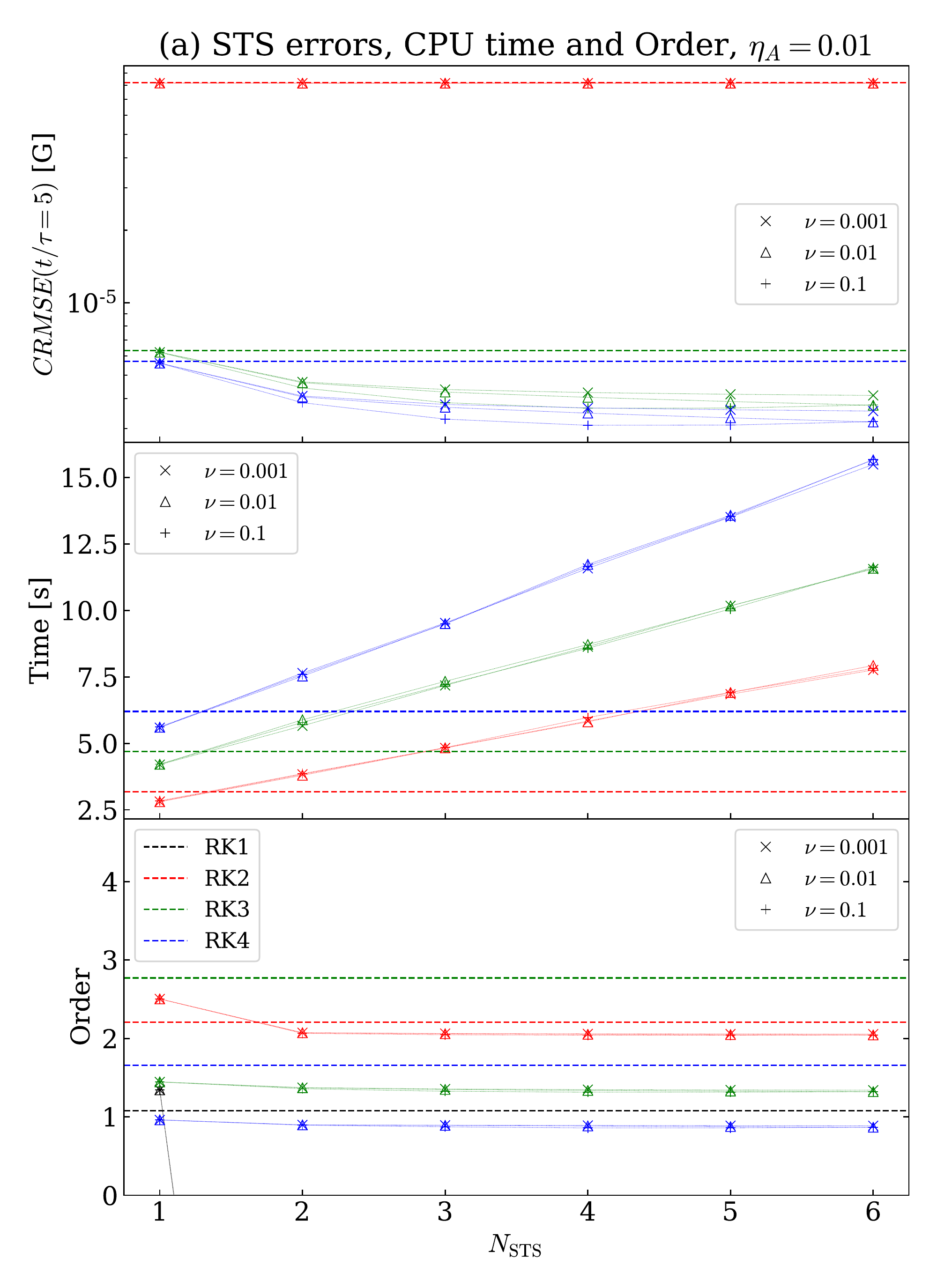}
    \includegraphics[width=0.49\textwidth]{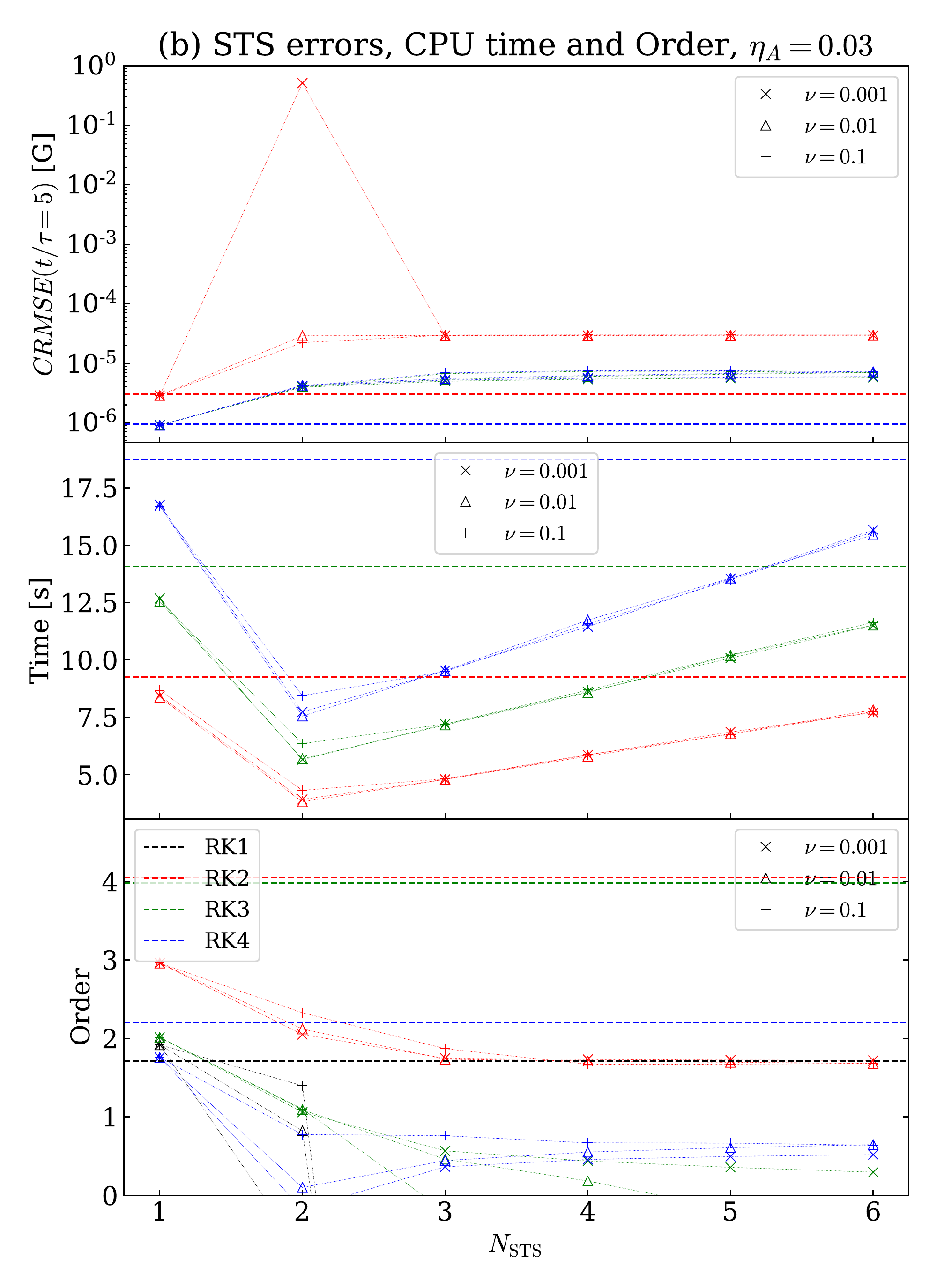}
	\caption{\footnotesize The upper panel shows the CRMSE values at $t=5\tau$ s for different values of the pair $(N_\mathrm{STS},\nu)$. The horizontal dashed lines corresponds to the RK scheme values. The symbols correspond to the STS scheme, each of them indicate a different value of $\nu$. The black, red, green and blue colours indicate first, second, third and fourth order accuracy respectively. The dotted lines linking the symbols are drawn just to make the plot more readable. The middle panel shows the CPU time spend by the schemes to reach the same point in the simulations. The color and symbol code is the same as the upper panel. The lower panel shows the order computed using the $L_1$-norm at $t_0=0.7\tau$, see the dotted vertical line in Fig.\ref{fig:ambitest}; again, the color and symbol code is the same as the upper panel. In subfigure (a), we have a case with low ambipolar diffusion where the first order results unstable and the errors for the STS scheme are very close to the RK values. In this case, we don't have acceleration because the system is out of the STS regime and the time increase linearly with $N_\mathrm{STS}$ as expected.}
	\label{fig:ambitest_errortime_1}
\end{figure*}

\begin{figure*}
	\centering
	\includegraphics[width=0.49\textwidth]{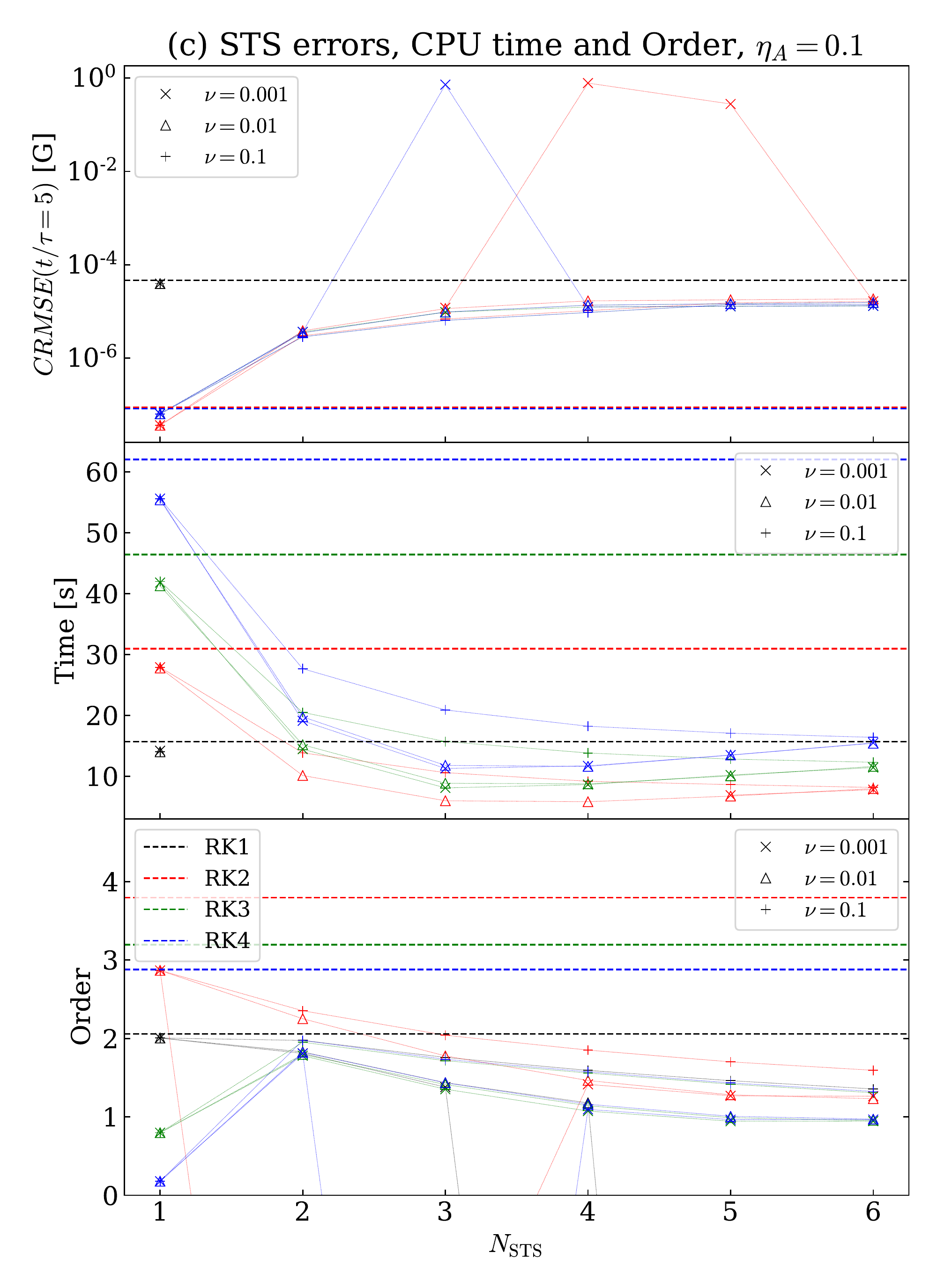}
	\includegraphics[width=0.49\textwidth]{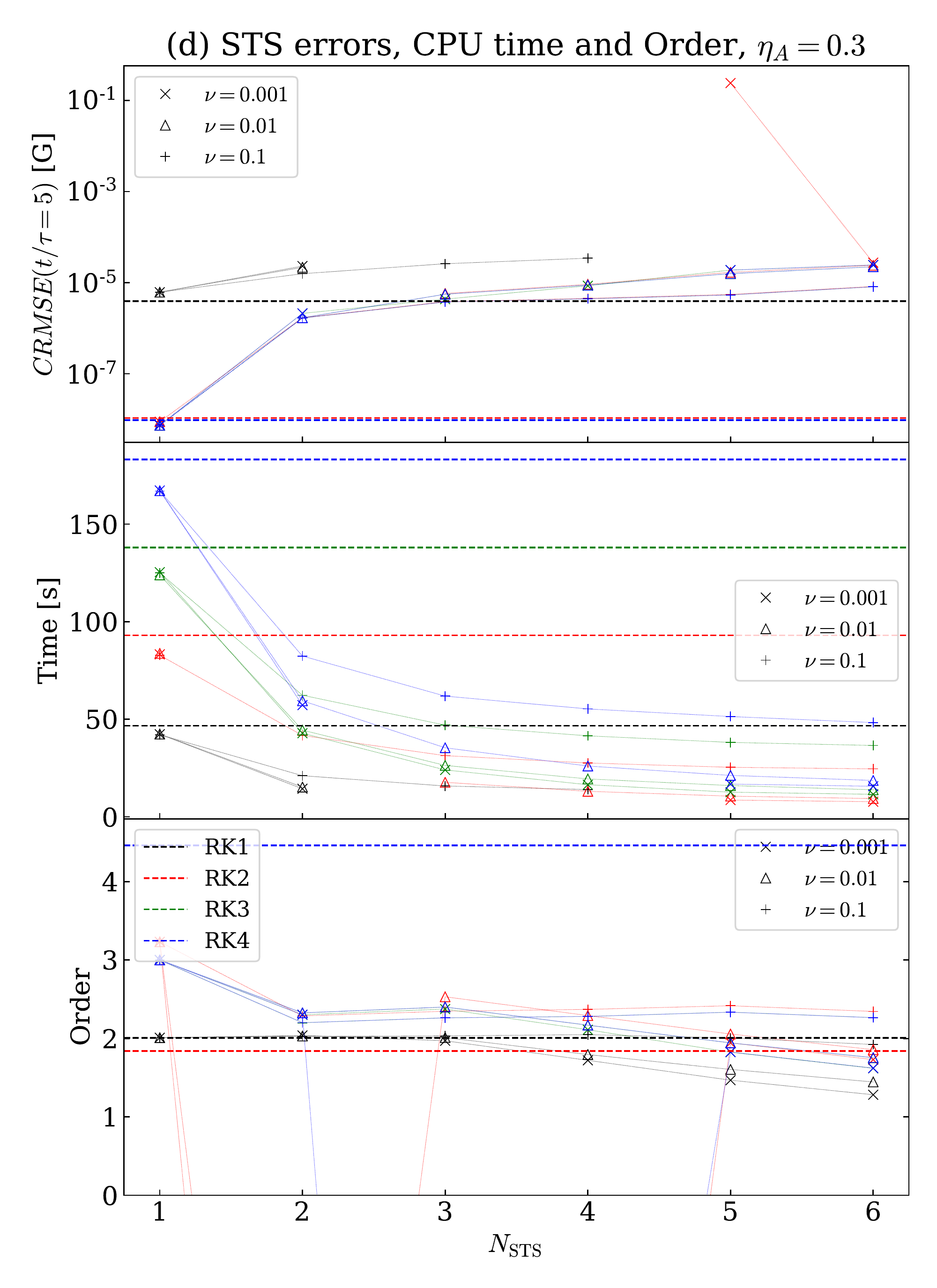}
	\caption{\footnotesize Same as Fig. \ref{fig:ambitest_errortime_1}, but showing a clear case inside the STS regime in subfigure (d). Here, the analytical solution is still well captured and the acceleration obtained is substantial. The errors, on the other hand, are limited, even though they are higher than in the RK1 case.}
	\label{fig:ambitest_errortime_2}
\end{figure*}

The results are shown in Figures \ref{fig:ambitest_errortime_1} and \ref{fig:ambitest_errortime_2} where in the upper panel of those figures we have plotted the CRMSE at a fixed time ($t=5\tau$, with $\tau=L/c_A$) as function of $(N_\mathrm{STS},\nu)$. The time spent by each simulation to reach the same point in time (also at $t=5\tau$) is plotted at the middle panel. At the lower panel we show the accuracy order measured at $t=0.7\tau$. We chose this time to avoid values of the solutions close to zero  (see vertical line marks in Fig. \ref{fig:ambitest}). All the tests where executed using a single node of the TeideHPC supercomputer (each TeideHPC node has 2 Intel Xeon E5-2670 processors, for a total of 16 cores per node).

In Figure \ref{fig:ambitest_errortime_1}a, upper panel, we observe how the accumulated error for STS at order 2 results very similar to the RK one. However, for higher orders, the error is slightly smaller than the one corresponding to the RK schemes (horizontal dashed-lines). We see how increasing $N_\mathrm{STS}$ (i.e., increasing the number of times the STS operator will be called) introduces small extra errors. On the other hand we see how the CPU time increases linearly with $N_\mathrm{STS}$ due to these extra calls. Such behaviour is expected because our system is outside of the STS regime due to the low ambipolar diffusion and in this case, performing STS involves unnecessary computational effort. In Fig. \ref{fig:ambitest_errortime_1}b the system gets into the STS regime only for $N_\mathrm{STS}=2$, so for the rest of values we see the same behaviour as before. In the other cases shown in Fig. \ref{fig:ambitest_errortime_2}, $\eta_A = 0.1$ or $0.3$, we have that the accumulated error increases over the error obtained for the RK1. However, by using the STS technique we obtain a speed-up of between two and eight times depending on the RK order chosen to compare. For $N_\mathrm{STS} > 4$ the decreasing time tendency starts to invert, see Figure \ref{fig:ambitest_errortime_2}c. This happens because in this particular test, the STS jump reaches the size of the MHD jump and the code starts to spend time with extra iterations of the STS scheme as in the previous cases.

Finally, following \citet{LeVeque:2002th}, to measure the accuracy order for conservation laws, we computed the $L_1$-norm against the analytical solution given by eq. (\ref{eq:ambi_sol}) as: 
\begin{equation}\label{eq:l1norm}
    L_1(t_0)= \frac{1}{N_p} \sum_{i=1}^{N_p}  | B_z(x_i,t_0)_\mathrm{Analytical} - B_z(x_i,t_0)_\mathrm{Numerical} | \coma
\end{equation}
where the $L_1$-norm is calculated at $t_0=0.7\tau$ (see vertical dotted line in Fig. \ref{fig:ambitest}) and $N_p$ is the number of points into the numerical domain. The results are in general consistent with the selected order of the schemes and limitation imposed by the Strang splitting. We notice nevertheless, that for this experiment, the rounding errors are affecting strongly the determination of the order because we quickly reach a precise solution, even when working with low orders. Thus, an increase of the order does not improve the solution, and therefore the value of the error calculated according to eq. (\ref{eq:l1norm}) is meaningless. This problem of the error evaluation can be seen clearly for the RK schemes with high ambipolar diffusion.

In the upper panel of Fig. \ref{fig:ambitest_errortime_2}c we observe how the CRMSE gets values close to one in some cases. These cases correspond to the lower value of $\nu$, and illustrate how the solution is entering into a unstable region, that is, the sub-steps $\tau_j$ are giving unstable values and the solution would not converge at the super-step. Checking the lower panel of the same figure, one can see how these points also correspond to the accuracy order values with no meaning. We conclude that one needs to be very careful in choosing the value of this damping parameter to avoid numerical instabilities. As we have seen, good values for this parameter can be obtained using eq. (\ref{eq:eff}). Another thing we notice is that for $N_\mathrm{STS} = 1$ the behaviour of the CRMSE, CPU time or order matches the RK scheme very well. This is because when $N_\mathrm{STS}$ is one, the damping parameter $\nu$ is zero and then, the STS acts as a MHD operator, (i.e., a RK scheme), and then only the Strang splitting is affecting the results. In this case in particular, the results can be seen in Fig.\ref{fig:ambitest} and verify that, despite the wrong values obtained for the accuracy order, see for example the lower panel of  Fig. \ref{fig:ambitest_errortime_2}c ($\eta_A=0.1$ m$^2$ s$^{-1}$), the numerical solution matches very well the analytic one.

For the STS we observe how by increasing $N_\mathrm{STS}$ or decreasing $\nu$ the CRMSE is increasing and the computation time and the accuracy order are decreasing. On the other hand, we see that by increasing the ambipolar diffusion coefficient the accuracy order tends to decrease slightly and both the CRMSE and the computational time tend to increase.

Despite the fact that STS is limited to second order due to the Strang splitting, we observe that setting the order of each operator individually at 3 or more, makes the whole system more stable. This way, we could capture better stiff problems and/or force the values of $(N_\mathrm{STS},\nu)$ to obtain a higher acceleration.

%%%%%%%%%%%%%%%%%%%%%%%%%%%%%%%%%%%%%%%%%%%%%%%%%%%%%%%%%
\subsection{HDS operator test}\label{sec:HDS_op}
%%%%%%%%%%%%%%%%%%%%%%%%%%%%%%%%%%%%%%%%%%%%%%%%%%%%%%%%%

In order to test the HDS operator we have considered an experiment with a plane-polarized Alfv\'en wave propagating into a partially ionised plasma under the presence of a guiding magnetic field as it is proposed by \citet{2012ApJ...750....6C}. For this experiment we have neglected all the induction terms except the Hall term.

The numerical set up for the test consists in a 2.5D periodical box with a constant hydrodynamic background and a pure Alfv\'en wave propagating along the $x$-axis. The transverse components are considered small compare with the guide field, $B_x$.

Under these conditions, the Hall coefficient, $\eta_H$ can be considered constant and the analytical solution given by equations (\ref{eq:hds_sol}a, \ref{eq:hds_sol}b, and \ref{eq:hds_sol}c), can be obtained after linearised the single-fluid equations. This solution describes the precession of the polarisation plane of an Alfv\'en wave with a wavenumber $k$ initially plane-polarised in the $y$-axis.
\begin{subequations}\label{eq:hds_sol}
    \begin{alignat}{1}
        B_x(x,t)= {} & B_0 \coma\\
        B_y(x,t)= {} & b\cos(\sigma t)\cos(kx)\cos(\omega t) \coma \\
        B_z(x,t)= {} & b\sin(\sigma t)\cos(kx)\cos(\omega t) \coma
    \end{alignat}
\end{subequations}
where $\omega$ is the angular frequency, $B_0$ is a constant uniform magnetic field, $b$ is a small perturbation and $\sigma$ is the precession rate of the polarisation plane.
\begin{equation}
    \sigma = \frac{1}{2}\eta_H k^2 \punto
\end{equation}
The relation between $k$ and $\omega$ is obtained from the  dispersion relation:
\begin{equation}
    \left(\frac{\omega}{k}\right)^2=c_A^2+\frac{1}{4}\eta_H^2 \punto
\end{equation}
For this experiment we selected physical parameter to match those of the solar photosphere: $\eta_H=10^6$ m$^2$ s$^{-1}$, $B_0=100$ G, $T=6000$ K, $b=0.1$ G, $p_{gas}=10^3$ N/m$^2$, $\rho=10^{-4}$ kg/m$^3$, and $L=100$ km so the domain covers one wavelength. For this set of parameters we have $k=2\pi \times 10^{-5}$ m, $\sigma=1.9\times10^{-3}$ Hz, $\omega=5.6\times10^{-2}$ Hz, and the rotational period $\tau = 2\pi\sigma^{-1}= 3183$ s. For the initial conditions we set:
\begin{subequations}\label{eq:hds_inicond}
    \begin{alignat}{1}
        B_x(t=0){} &=B_0 \coma \\
        B_y(t=0){} &=b\cos(kx) \coma \\
        B_z(t=0){} &=0 \punto
    \end{alignat}
\end{subequations}

\begin{figure}[t]
    \resizebox{\hsize}{!}{\includegraphics{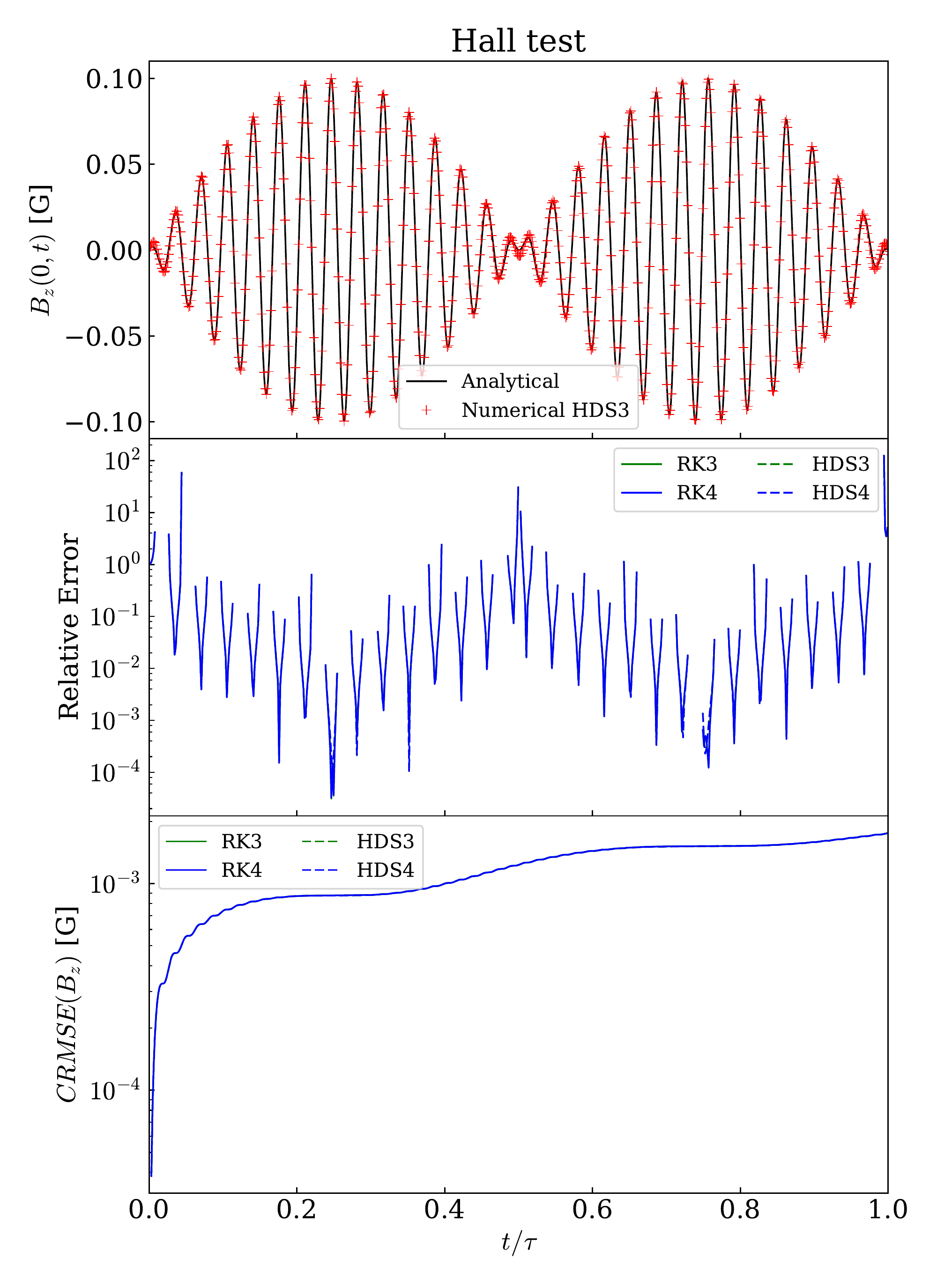}}
    \caption{\footnotesize Hall test, showing in the upper panel the time evolution for the numerical and analytical solutions of the $B_z$ components of the magnetic field fluctuations due to the precession of the plane-polarised Alfv\'en wave in a fixed spacial point ($x=0$). The solid line represents the analytical solution and the red crosses are the numerical solution. In the middle panel it is plot the relative error of the numerical solution respect the analytical solution for both the RK and the HDS schemes, solid lines and dashed lines respectively. The lower panel shows the CRMSE in time for the $B_z$ component. It can be seen how the two schemes have an almost identical behaviour in time and with the order. This it is expected because the HDS operator works as a RK operator. Also this shows us how well performs the Strang splitting.}
    \label{fig:halltest}
\end{figure}
In the upper panel of Figure (\ref{fig:halltest}) we can see the result of the simulations over the analytical solution given by equations (\ref{eq:hds_sol}). In the middle panel we can see how the relative error is higher near the nodes and, in the lower panel, how the CRMSE increase in time as it is expected. We notice that simulations at first and second order are not plotted because they were not stable for the chosen CFL. However, at third and fourth order both schemes captures very well the analytical solution and they are very similar between them. Reducing the CFL slightly, the second order becomes stable.

%%%%%%%%%%%%%%%%%%%%%%%%%%%%%%%%%%%%%%%%%%%%%%%%%%%%%%%%%
\subsection{MHDSTS scheme test}\label{sec:MHDSTS_scheme}
%%%%%%%%%%%%%%%%%%%%%%%%%%%%%%%%%%%%%%%%%%%%%%%%%%%%%%%%%

Finally, this section shows the results of testing simultaneously the MHD, HDS, and STS operators considering the effects of the Ohmic and ambipolar diffusions and the Hall term. For that, we used a shock-tube test with a magnetic precursor, or C-shocks \citep{1980ApJ...241.1021D}, under two different regimes depending of what term is dominating over the other non-ideal terms: a) Hall dominated and b) ambipolar dominated, see \citet{Falle:2003er,2006MNRAS.366.1329O,2007MNRAS.376.1648O}.

To obtain the steady-state solution we consider our system isothermal with all the three diffusivity coefficients ($\eta$, $\eta_A$, and $\eta_H$) constants and no battery term. We also set all the time derivatives to zero, this is equivalent to transform the equations to a frame of reference where the shock is steady.
With this, our MHD equations can be written as
\begin{subequations}
    \begin{align}
        &\rho \mathrm{v}_z=C_0 \coma\\
        &\rho \mathrm{v}_x \mathrm{v}_z - \frac{1}{\mu_0}B_x B_z = C_x \coma \\
        &\rho \mathrm{v}_y \mathrm{v}_z - \frac{1}{\mu_0}B_y B_z = C_y \coma \\
        &\rho \mathrm{v}_z^2+c_s^2\rho+\frac{1}{2\mu_0}B^2-\frac{1}{\mu_0} B_z^2=C_z\coma
    \end{align}
\end{subequations}
with the induction equation as
\begin{subequations}
    \begin{align}
    \begin{split}
    \left[\eta + \frac{\eta_A}{|B|^2}(B_x^2+B_z^2)\right] \frac{\partial B_x}{\partial z} +{}&
    \left[\frac{\eta_H}{|B|}B_z + \frac{\eta_A}{|B|^2}B_xB_y\right] \frac{\partial B_y}{\partial z} + {}\\& +
    \mathrm{v}_xB_z-\mathrm{v}_zB_x = C_1\coma
    \end{split}
    \allowdisplaybreaks \\
    \begin{split}
    \left[\frac{\eta_A}{|B|^2}B_xB_y - \frac{\eta_H}{|B|}B_z\right] \frac{\partial B_x}{\partial z} +{}&
    \left[\eta + \frac{\eta_A}{|B|^2}(B_y^2+B_z^2)\right] \frac{\partial B_y}{\partial z} + {}\\& +
    \mathrm{v}_yB_z-\mathrm{v}_zB_x = C_2\coma
    \end{split}
    \end{align}
\end{subequations}
where $C_0$, $C_x$, $C_y$, and $C_z$ are constants in time. $C_1$ and $C_2$ are constants of integration obtained from applying the pre-shock boundary condition.

The initial states for the pre-shock and post-shock plasma were obtained from the jump conditions of a front shock propagating into a magnetised fluid. The parameter for this initial state are given in Table \ref{tab:ini_cond}.

Given the density and velocity at the post-shock side we can determine the magnetic field from the induction equation at each spatial position and then use the other four algebraical equations to obtain the rest of variables. To solve this ordinary differential system of equations with non-constant coefficients, we use a fourth order RK scheme implemented into a different code.

\begin{table}[ht]
    \caption{\footnotesize Parameters for the MHDSTS test. The subscript 1 corresponds to the pre-shock side and the subscript 2 with the post-shock region. All the parameters are in the S.I.}
    \label{tab:ini_cond}
    \tiny
    \centering
    \begin{tabular}{c|lll}
    \hline
    Ini. & $\rho_1 = 1$ & $\textbf{v}_1=(0, 0, -1.7)$ & $\textbf{B}_1=(0.6 , 0, 1)\mu_0^{1/2}$  \\
    Cond. &$\rho_2 = 1.8$ & $\textbf{v}_2=(-0.6, 0, -1)$ & $\textbf{B}_2=(1.7, 0, 1)\mu_0^{1/2}$  \\
    \hline
    Case & $\eta = \mu_0$ & $\eta_A=25\mu_0/|\textbf{B}|^2$ & $\eta_H=500\mu_0/|\textbf{B}|$ \\
    (a) & $\nu=0$ & $N_\textrm{STS}=1$ & $N_\textrm{HDS}=6$ \\
    \hline
    Case & $\eta = \mu_0$ & $\eta_A=300\mu_0/|\textbf{B}|^2$ & $\eta_H=750\mu_0/|\textbf{B}|$ \\
    (b)  & $\nu=0.15$ & $N_\textrm{STS}=2$ & $N_\textrm{HDS}=8$ \\
    \hline
    \end{tabular}
\end{table}

\begin{figure}[t]
    \resizebox{\hsize}{!}{\includegraphics{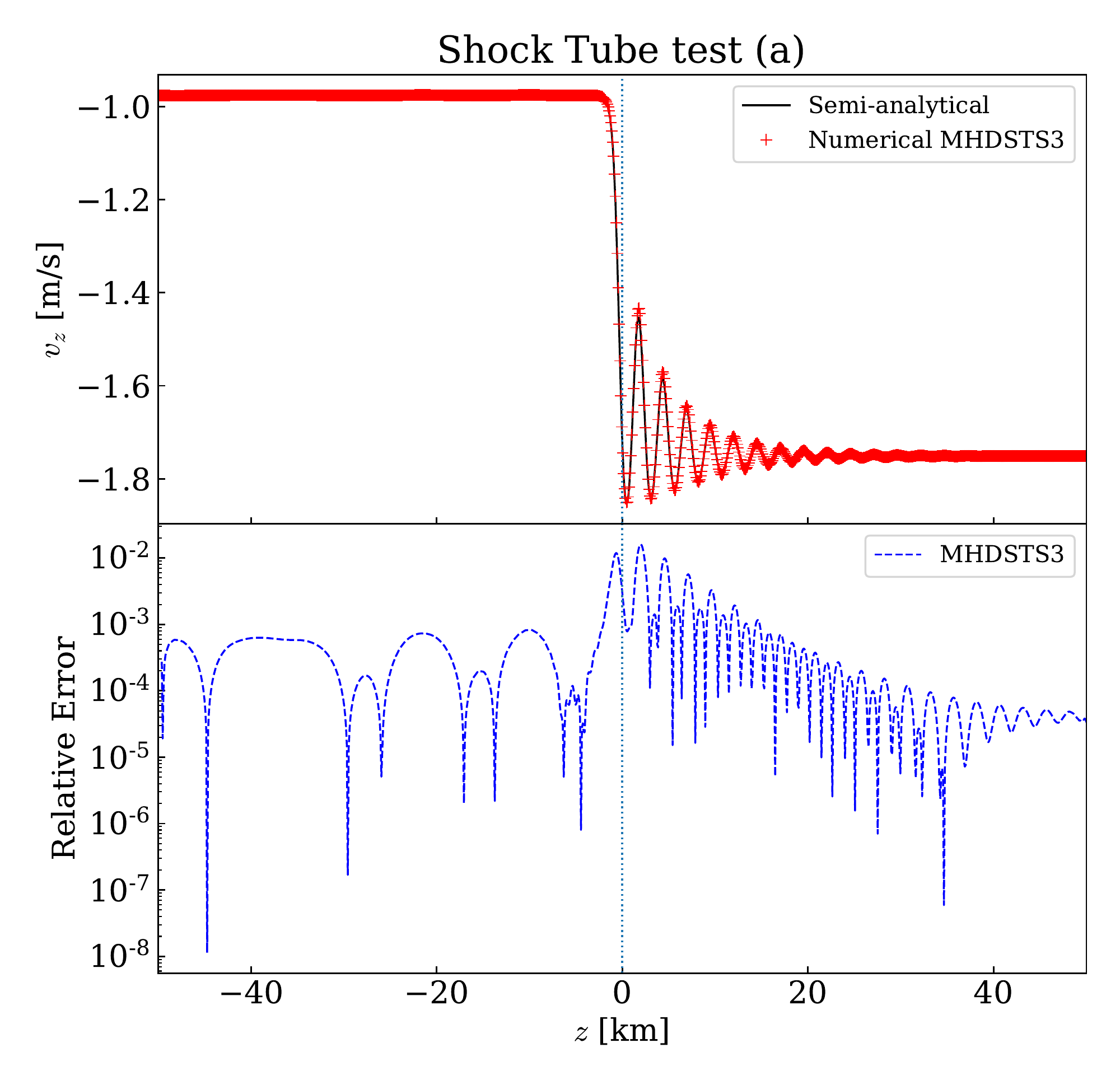}}
    \caption{\footnotesize Hall dominated shock tube test. The upper panel shows the steady solution for the z-component of the velocity field. The left side corresponds to the post-shock side. The lower panel shows the relative error. The order obtained for this experiment is $\sim 2.1$.}
    \label{fig:shocktube_hall}
\end{figure}

\begin{figure}[t]
    \resizebox{\hsize}{!}{\includegraphics{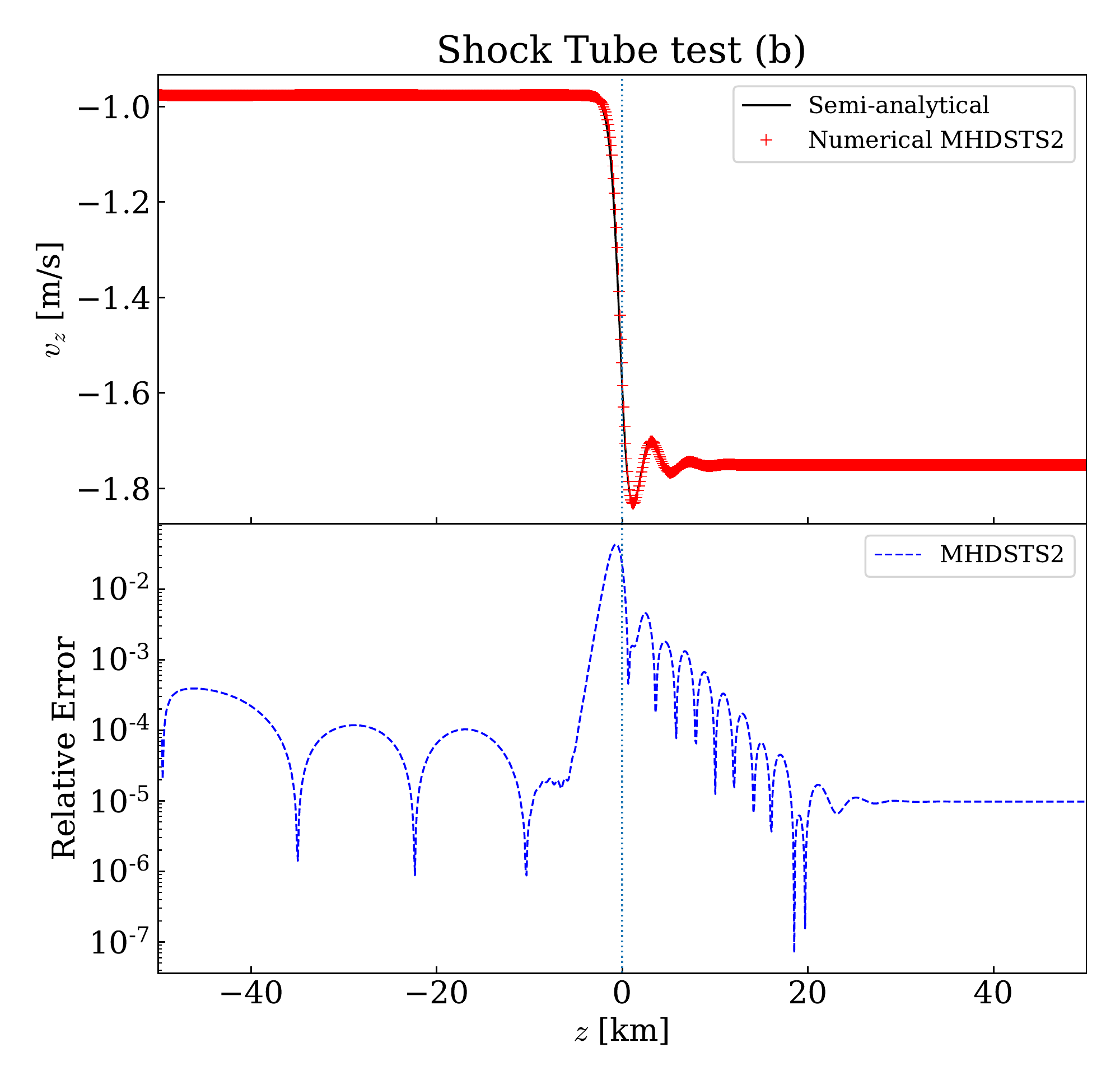}}
    \caption{\footnotesize Shock tube test with high ambipolar diffusion and comparable with the Hall term. The upper panel shows the steady solution for the z-component of the velocity field. The left side corresponds to the post-shock side. The lower panel shows the relative error. The order obtain from this test is $\sim 1.6$ a slightly lower than the experiment (a) due to the STS contribution as it was commented in Subsection \ref{sec:STS_op}.}
    \label{fig:shocktube_ambihall}
\end{figure}

With these parameters, and avoiding the use of the artificial diffusion terms, the calculation is unstable for the RK scheme at all orders due to the high frequency oscillation (whistler waves) at the shock front introduced by the Hall term. The only way of solving this problem with the RK scheme without artificial diffusion is by forcing a reduced CFL condition and increasing the filtering cadence. Then, as a side effect we will spend more CPU time and tend to over-smooth the solution. The whistler waves can be seen clearly at the upper panel of Fig. \ref{fig:shocktube_hall}, where we can also see how the analytical solution is very well captured by the MHDSTS scheme. In this case, if we choose order 2 without artificial diffusion, the experiment results unstable but choosing order 3 the experiment is stabilised due to the slightly higher numerical diffusivity introduced by the extra RK-step to reach order 3. In the lower panel of the same figure we can see the relative error between both solutions. For this experiment, as we have obtained a semi-analytical solution, it is easy to compute the accuracy order by changing the mesh resolution. In this case the obtained order is $\sim 2.1$. In Fig. \ref{fig:shocktube_ambihall}, we show a second experiment with a higher ambipolar diffusion to activate properly the STS operator but keeping also a high Hall term contribution. We see again the good agreement between both solutions and also how by increasing the ambipolar diffusion enough, the selected global order of the code needed to get the solution is lower. In this case, the calculated accuracy order is $\sim 1.6$, slightly smaller than the case (a) due to the action of the STS scheme.

%%%%%%%%%%%%%%%%%%%%%%%%%%%%%%%%%%%%
\section{Conclusions}\label{sec:sum}
%%%%%%%%%%%%%%%%%%%%%%%%%%%%%%%%%%%%
In this paper we have seen that STS and HDS are numerical techniques easy to implement and also that they can work together using the Strang operator splitting formalism forming what we have called MHDSTS scheme. We also show how it is possible to increase the temporal order of accuracy using a RK formalism as a wrapper of each operator. However, MHDSTS is limited to second order by the Strang splitting, but setting the order of each operator individually at 3 or more makes the whole system more stable.

So far, all the tests performed have shown a good agreement with the analytical solution. This make us confident enough to use this new numerical scheme with \mancha code in a production mode to investigate the effects of those diffusion terms with stiff problems.

We also check that the STS technique can speed up problematic simulations by overcoming the CFL condition imposed by the parabolic term corresponding to the ambipolar term. But as an alert note, we have to be aware that we must choose carefully the pair $\nu$ and $N_\textrm{STS}$ because, as \citet{Alexiades:1996vj} mentions, and we show with a numerical test, the smaller $\nu$ and/or bigger $N_\textrm{STS}$ means bigger the acceleration but also bigger errors and the increase of numerical instabilities, as well as a slight decrease of the accuracy order.

On the other hand, we also check that the HDS technique solves the problem with complex eigenvalues introduced by Hall term being more stable than our standard  Runge-Kutta scheme and avoiding that our $dt_\mathrm{Hall}$ becomes very small, specially when the Hall term is dominating. 

Working with the MHDSTS scheme, we have that the HDS time step is improved when the ambipolar or Ohmic diffusion are considered.

Finally, due to the whole scheme being explicit, even when HDS behaves as an implicit-like scheme, it is straightforward to use the other already implemented capabilities of the \mancha code such as parallel scalability, the PML boundary condition, the AMR capability or the radiative transfer module.

\begin{acknowledgements}
     This work is supported by the predoctoral training grants in Centres/Units of Excellence "Severo Ochoa". We also wish to acknowledge the contribution of Teide High-Performance Computing facilities to the results of this research. TeideHPC facilities are provided by the Instituto Tecnol\'ogico y de Energ\'ias Renovables (ITER, SA). URL: http://teidehpc.iter.es
\end{acknowledgements}

% WARNING
%-------------------------------------------------------------------
% Please note that we have included the references to the file aa.dem in
% order to compile it, but we ask you to:
%
% - use BibTeX with the regular commands:
\bibliographystyle{aa} % style aa.bst
\bibliography{mhdsts} % your references Yourfile.bib
%\bibliography{solar}
%
% - join the .bib files when you upload your source files
%-------------------------------------------------------------------

\end{document}